\documentclass[aps,prd,floatfix,12pt]{revtex4}
\usepackage{epsfig}
\usepackage{bm}
\usepackage{dcolumn}
\usepackage{latexsym}

\begin{document}
   
\preprint{\rightline{ANL-HEP-PR-10-45}}
   
\title{The chiral phase transition for lattice QCD with 2 colour-sextet quarks}

\author{J.~B.~Kogut}
\affiliation{Department of Energy, Division of High Energy Physics, Washington,
DC 20585, USA}
 \author{\vspace{-0.2in}{\it and}}
\affiliation{Dept. of Physics -- TQHN, Univ. of Maryland, 82 Regents Dr.,
College Park, MD 20742, USA}
\author{D.~K.~Sinclair}
\affiliation{HEP Division, Argonne National Laboratory, 9700 South Cass Avenue,
Argonne, IL 60439, USA}

\begin{abstract}
QCD with 2 flavours of massless colour-sextet quarks is studied as a possible 
walking-Technicolor candidate. We simulate the lattice version of this model at
finite temperatures near to the chiral-symmetry restoration transition, to 
determine whether it is indeed a walking theory (QCD-like with a running 
coupling which evolves slowly over an appreciable range of length scales) or if
it has an infrared fixed point, making it a conformal field theory. The lattice
spacing at this transition is decreased towards zero by increasing the number 
$N_t$ of lattice sites in the temporal direction. Our simulations are performed 
at $N_t=4,6,8,12$, on lattices with spatial extent much larger than the temporal
extent. A range of small fermion masses is chosen to make predictions for the
chiral (zero mass) limit. We find that the bare lattice coupling does decrease 
as the lattice spacing is decreased. However, it decreases more slowly than
would be predicted by asymptotic freedom. We discuss whether this means that 
the coupling is approaching a finite value as lattice $N_t$ is increased -- the
conformal option, or if the apparent disagreement with the scaling predicted by
asymptotic freedom is because the lattice coupling is a poor expansion 
parameter, and the theory walks. Currently, evidence favours QCD with 2
colour-sextet quarks being a conformal field theory. Other potential sources
of disagreement with the walking hypothesis are also discussed.

    We also report an estimate of the position of the deconfinement transition
for $N_t=12$, needed for choosing parameters for zero-temperature simulations.
\end{abstract}

\maketitle

\section{Introduction}

The LHC at CERN is currently probing the Higgs sector of the Standard Model of
high-energy physics. This sector is the least well understood part of the
standard model, and the least satisfactory from a theoretical standpoint. Thus
the study of extensions of the standard model with a more aesthetically
compelling Higgs sector is timely. The observation of a light ($\approx
125$~GeV) Higgs-like excitation at ATLAS and CMS, with properties consistent
with the standard-model Higgs, puts constraints on any such model. We are
especially interested in those models where the Higgs sector is
strongly-coupled and the Higgs boson is composite.

We are interested in QCD-like models -- non-Abelian gauge theories with massless
fermions and spontaneously broken chiral symmetry -- where the pion-like 
Goldstone bosons play the role of the Higgs field, giving mass to the $W^{\pm}$ 
and $Z$ weak vector bosons through the Higgs mechanism. Here the Higgs boson is
the remnant radial excitation. Such theories are called Technicolor models
\cite{Weinberg:1979bn,Susskind:1978ms}. 
Technicolor models, which are simply QCD scaled up so that 
$f_\pi \approx 246$~GeV rather than $f_\pi \approx 93$~MeV of regular QCD, are
not phenomenologically viable. It has been suggested that Technicolor theories
where the fermion content is such that the running gauge-coupling evolves
very slowly over an appreciable range of mass scales, described as `walking'
rather than running, might be capable of overcoming such difficulties. Such
theories are referred to as Walking Technicolor models
\cite{Holdom:1981rm,Yamawaki:1985zg,Akiba:1985rr,Appelquist:1986an}. 
Because of their nature,
the non-perturbative properties of such models are amenable to study using the
simulation methods developed for Lattice QCD. It is such theories that we are
interested in simulating.

Candidate walking gauge theories typically have 2-loop $\beta$-functions with a
second, non-trivial, zero. If this behaviour remains true to all orders, this 
non-trivial fixed point controls the infrared properties of the theory, which
is therefore a conformal field theory with a continuous spectrum. On the 
otherhand, if the running coupling becomes so large that a chiral condensate
forms before this would-be IR fixed point is reached, the theory is QCD-like
with spontaneously broken chiral symmetry and Goldstone bosons separated by a
mass gap from the rest of the spectrum. Because of the proximity of the would-be
fixed point, a region where the coupling walks would be expected.

We have concentrated our efforts on techni-QCD with 2 flavours of massless 
Technicolor-sextet techni-quarks. Since this is identical to QCD with 2 
colour-sextet quarks scaled so that $f_\pi \approx 246$~GeV, we will omit the
prefix `techni' from here on. This theory has been identified as a potential
walking-Technicolor candidate (see for example \cite{Dietrich:2006cm} and
references therein). It is asymptotically free and its 2-loop 
$\beta$-function does have a non-trivial zero far enough from $g^2=0$ for the
coupling to become sufficiently large for there to be a chance that chiral
symmetry breaks before it is reached. If it is indeed QCD-like, it has 3
Goldstone bosons, the correct number to give masses to the $W$s and $Z$, with
none left over. In this sense, it is minimal.

Other groups have studied/are studying this model using lattice techniques. The
main contributors are Degrand, Shamir and Svetitsky
\cite{Shamir:2008pb,DeGrand:2008kx,DeGrand:2009hu,DeGrand:2010na,%
DeGrand:2012yq,DeGrand:2013uha}
and the Lattice Higgs Collaboration
\cite{Fodor:2009ar,Fodor:2011tw,Fodor:2012ty,Fodor:2012uw,Fodor:2014pqa,%
Fodor:2015vwa,Fodor:2015eea,Fodor:2015zna},
In addition we should mention some recent work by A.~Hasenfratz and her 
collaborators \cite{Hasenfratz:2015ssa}.
With the exception of one early paper by Degrand, Shamir and Svetitsky, 
\cite{DeGrand:2008kx}
these have concentrated on the zero-temperature properties of this model. We
study lattice QCD with 2 colour-sextet quarks at finite temperature. Our goal
is to determine if the evolution of the coupling as lattice spacing 
$a \rightarrow 0$ is described by asymptotic freedom, and that chiral symmetry 
remains broken in this limit. Assuming that the chiral-symmetry-restoration
transition is indeed a finite-temperature transition, increasing the temporal
extent of the lattice, $N_t$ in lattice units, with the spatial extent 
$N_s >> N_t$, and with temperature $T$ fixed at the chiral phase transition 
temperature $T_\chi$, takes $a=1/N_tT_\chi$ towards zero. Thus the running
coupling at this temperature, $g_\chi(a)$, should approach zero as 
$N_t \rightarrow \infty$ in a manner determined by the perturbative
$\beta$-function. If, on the otherhand, the theory is conformal, $g_\chi$ will
approach a finite limit as $N_t \rightarrow \infty$, characterising a bulk
transition. Similar arguments should apply to the deconfinement transition at
$g_d$. However, as we have determined, deconfinement occurs at a much stronger
coupling than chiral-symmetry restoration. For the $N_t$ values we have
considered ($N_t \le 12$), $g_d$ is too large for its evolution to be
controlled by the perturbative $\beta$-function.

Recent use of this method, to search for the lower bound (in $N_f$) of the
conformal window for QCD with many fundamental quarks
\cite{Deuzeman:2008sc,Deuzeman:2009mh,Schaich:2012fr,Schaich:2015psa}
have shown it to complement step-scaling methods. The reader should consult
the references in these papers for the history of such studies. Although these
methods are, in principle, straight forward, these recent papers indicate that
they are not so easy to implement in practice.

Our earlier studies at $N_t=4,6,8$ \cite{Kogut:2010cz,Kogut:2011ty}
were consistent with the evolution of 
$g_\chi$ between $N_t=6$ and $N_t=8$ being described by the 2-loop 
$\beta$-function. We have extended our simulations to $N_t=12$. In addition, 
we have covered the neighbourhood of the chiral transition for $N_t=6$ and 
$N_t=8$ with more closely spaced values of $\beta=6/g^2$ to determine $g_\chi$
more precisely. In addition, we have determined the position of the
deconfinement transition for $N_t=12$ for one mass value. Preliminary versions
of the results presented in this paper have been presented at lattice
conferences \cite{Sinclair:2012fa,Sinclair:2014cga}.

While the observed change in $\beta_\chi=6/g_\chi^2$ between $N_t=6$ and 
$N_t=8$ is consistent with that predicted using the 2-loop $\beta$-function,
that between $N_t=8$ and $N_t=12$ is only about half the predicted value. At
face value, this suggests that $\beta_\chi$ could be approaching a finite limit,
which would mean that this theory is conformal. However, we need to be cautious,
since we are studying the evolution of the bare lattice coupling, which is 
known to be a poor choice of expansion parameters \cite{Lepage:1992xa}. 
In addition, we are using
unimproved staggered fermions for which perturbation expansions in terms of the
bare lattice coupling are particularly poorly behaved, because of the 
`taste'-breaking tadpoles \cite{Patel:1992vu,Golterman:1998jj}. 
We discuss this and other potential sources of systematic errors in our
approach, later in this paper. Our results make it important to perform further
studies of lattice QCD with 3 massless, colour-sextet quarks, to determine if
it approaches its expected asymptotic behaviour at $N_t=12$, which would be
qualitatively different from that observed for 2 flavours.

In section~2 we discuss our methods of simulation and analysis. Section~3 gives
the results of our simulations at $N_t=6,8$ and $12$ near the chiral transition.
We compare these results with perturbative predictions in section~4, and discuss
improvements. In section~5 we analyse simulations at $N_t=12$ in the 
neighbourhood of the deconfinement transition. In section~6 we discuss our
results, try to draw conclusions, and indicate directions for future studies.

\section{Methods}

This section largely repeats discussions given in earlier publications
\cite{Kogut:2010cz,Kogut:2011ty}, and is
included here for completeness. We use the simple Wilson plaquette action for
the gauge fields:
\begin{equation}
S_g=\beta \sum_\Box \left[1-\frac{1}{3}{\rm Re}({\rm Tr}UUUU)\right].
\end{equation}
Here the gauge fields $U$ on the links are in the fundamental representation of
$SU(3)_{\it colour}$. We use the unimproved staggered-fermion action for the
quarks: 
\begin{equation}
S_f=\sum_{sites}\left[\sum_{f=1}^{N_f/4}\psi_f^\dagger[D\!\!\!\!/+m]\psi_f
\right],
\end{equation}
where $D\!\!\!\!/ = \sum_\mu \eta_\mu D_\mu$ with 
\begin{equation}
D_\mu \psi(x) = \frac{1}{2}[U^{(6)}_\mu(x)\psi(x+\hat{\mu})-
                            U^{(6)\dagger}_\mu(x-\hat{\mu})\psi(x-\hat{\mu})],
\end{equation}
where $U^{(6)}$ is the sextet representation of $U$, i.e. the symmetric part of
the tensor product $U \otimes U$. Reasons for this choice have been discussed 
in our earlier publications. When $N_f$ is not a multiple of $4$ we use the 
fermion action:
\begin{equation}
S_f=\sum_{sites}\chi^\dagger\{[D\!\!\!\!/+m][-D\!\!\!\!/+m]\}^{N_f/8}\chi.
\end{equation}
The operator which is raised to a fractional power is positive definite and
we choose its positive-definite root. This yields a well-defined operator.
We use the RHMC method for our simulations \cite{Clark:2006wp}, where the
required powers of the quadratic Dirac operator are replaced by diagonal
rational approximations, to the desired precision. By applying a global
Metropolis accept/reject step at the end of each trajectory, errors due to the
discretization of molecular-dynamics `time' are removed.

The canonical partition function for a field theory at finite temperature $T$ is
realized by evaluating the functional integral for Euclidean time where the
time is restricted to an interval $1/T$ with periodic boundary conditions on
the boson fields and antiperiodic boundary conditions on the fermion fields. 
Space is kept infinite. On a lattice of lattice spacing $a$, this means using
a lattice of temporal extent $N_t$ in lattice units where $N_t a=1/T$. The
spatial extent of the lattice $N_s >> N_t$. For lattice QCD with sextet quarks,
if deconfinement and chiral-symmetry restoration are finite-temperature
transitions, then the associated temperatures at which they occur, $T_d$ and
$T_\chi$ respectively, should be fixed, independent of $a$, for $a$ small 
enough. Thus measuring the couplings at either of these transitions as $N_t$
is varied, gives $g(a)$ for a sequence of $a$s which approaches zero as
$N_t \rightarrow \infty$. As it turns out $\beta_d$ and hence $g_d$ lies in the
strong-coupling domain, outside the regime where perturbation theory is likely
to be valid, for $N_t=4,6,8,12$ and any other $N_t$ which we are likely to 
consider in the near future. We therefore concentrate our efforts on the chiral
transition, which occurs at much weaker couplings. If, on the other hand, QCD
with 2 colour-sextet quarks is conformal, the chiral transition would be a bulk
transition. In this case $g_\chi$ would approach a non-zero limit for large
$N_t$, and the whole region of broken chiral symmetry would be a lattice
artifact, disconnected from the conformal field theory at weaker coupling.

If QCD with 2 colour-sextet quarks is QCD-like, the approach of $g_\chi$ to
zero is described by asymptotic freedom expressed in terms of the 
$\beta$-function. Through 2 loops this is given by:
\begin{equation}
\beta(g) = -b_1 g^3 - b_2 g^5.
\end{equation}
Expressed in terms of $\beta=6/g^2$, the evolution of the coupling when the
lattice spacing is scaled by $\lambda$ is given by
\begin{equation}
\Delta\beta(\beta) = \beta(a) - \beta(\lambda a)
                   = (12b_1 + 72b_2/\beta)\ln(\lambda) + {\cal O}(1/\beta^2),
\label{eqn:deltabeta}
\end{equation}
where for $N_f$ flavours of colour-sextet quarks:
\begin{eqnarray}
b_1 &=& \left(11 - \frac{10}{3}N_f\right)/16\pi^2 \nonumber \\
b_2 &=& \left(102 - \frac{250}{3}N_f\right)/(16\pi^2)^2 .
\end{eqnarray}

The chiral transition occurs at that value of $\beta$ ($\beta_\chi$) at which
the chiral symmetry is restored and beyond which the chiral condensate
$\langle\bar{\psi}\psi\rangle$ vanishes, for massless quarks. Of course, in
lattice simulations, we need to run at (small but) finite mass, and extrapolate
to zero quark mass. It is not, however, practical to run at masses small
enough for the condensate to accurately determine $\beta_\chi$ directly. We 
therefore estimate $\beta_\chi$ from the peaks in the chiral susceptibilities,
or rather in the disconnected part of the chiral susceptibilities. This is
given by:
\begin{equation}
\chi_{\bar{\psi}\psi} = V\left[\langle(\bar{\psi}\psi)^2\rangle
                      -        \langle\bar{\psi}\psi\rangle^2\right]
\label{eqn:chi}
\end{equation}
where the $\langle\rangle$ indicates an average over the ensemble of gauge
configurations and $V$ is the space-time volume of the lattice. Since we use
stochastic estimators for $\bar{\psi}\psi$, we need at least 2 estimators per
configuration. The first term must include only contributions which are 
off-diagonal in the noise, to obtain an unbiased estimator. We, in fact, use
5 stochastic estimators at the end of each trajectory giving 10 estimates for
$\chi_{\bar{\psi}\psi}$ per configuration.

\section{Simulations of QCD with 2 flavours of colour-sextet quarks at
         $N_f=6,8,12$}

Here we describe our simulations with 2 colour-sextet quarks on lattices with
$N_f=6,8$ and $12$. For $N_f=6$ and $8$ we have extended the simulations of our
earlier papers, where we simulated at $\beta$ spacings of $0.1$, through the
chiral transition region. For the lowest mass ($m=0.005$) at $N_f=6$, we have
covered the vicinity of the chiral transition at $\beta$ spacings of $0.02$,
and with increased statistics. At $N_t=8$ we have also covered the vicinity of
the chiral transition with $\beta$ spacings of $0.02$ for all masses, including
a new smaller mass ($m=0.0025$). We have performed new high-statistics
simulations at $N_t=12$, covering the region of the chiral transition with
$\beta$ spacings of $0.02$, for all masses. Preliminary results of these new
simulations have been presented at Lattice 2011, Lattice 2012, Lattice 2013 and
Lattice 2014. In each case, we simulate using the RHMC algorithm with trajectory
length 1. Most of our simulations have been performed on lattices with 
$N_s=2N_t$.

\subsection{$N_t=6$}

\begin{figure}[htb]
\epsfxsize=6.0in
\centerline{\epsffile{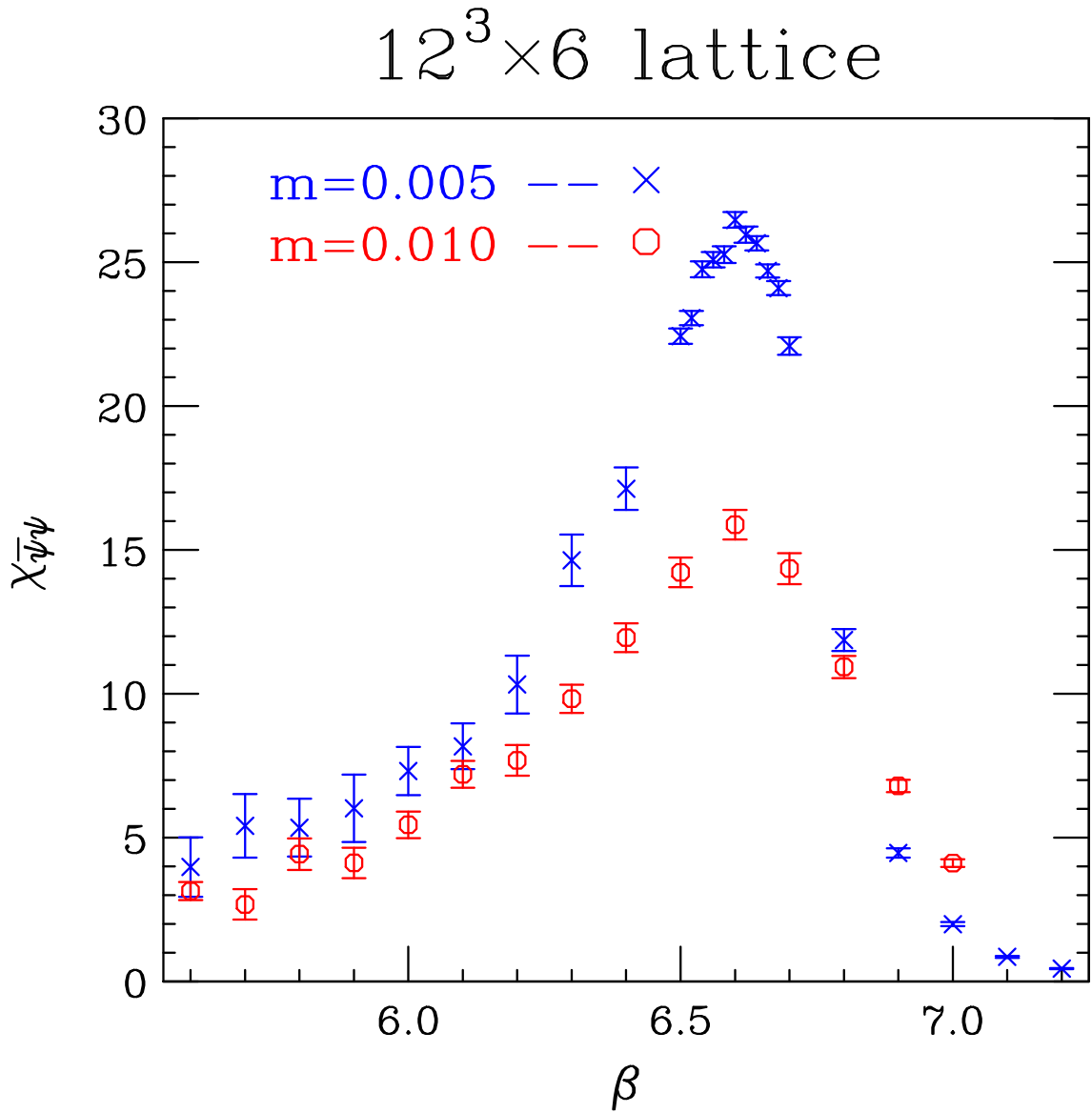}}
\caption{Chiral susceptibilities on a $12^3 \times 6$ lattice, with $N_f=2$.}
\label{fig:chi6}
\end{figure}

Our simulations at $N_t=6$ are performed on $12^3 \times 6$ lattices with quark
masses $m=0.005$, $m=0.01$ and $m=0.02$. To more accurately pinpoint the peak of
the chiral susceptibility at the lowest mass ($m=0.005$) we have covered the
region in the neighbourhood of the chiral transition $6.5 \le \beta \le 6.7$ at
$\beta$ spacings of $0.02$. At each of these $\beta$s we have performed runs of
100,000 trajectories. Outside of this interval, and that near the deconfinement
regime, we employ $\beta$ spacings of $0.1$ and 10,000 trajectories per $\beta$.
The chiral susceptibilities (equation~\ref{eqn:chi}) from these new runs and
those for $m=0.01$ from our earlier work are plotted in figure~\ref{fig:chi6}.

\begin{figure}[htb]
\epsfxsize=6.0in
\centerline{\epsffile{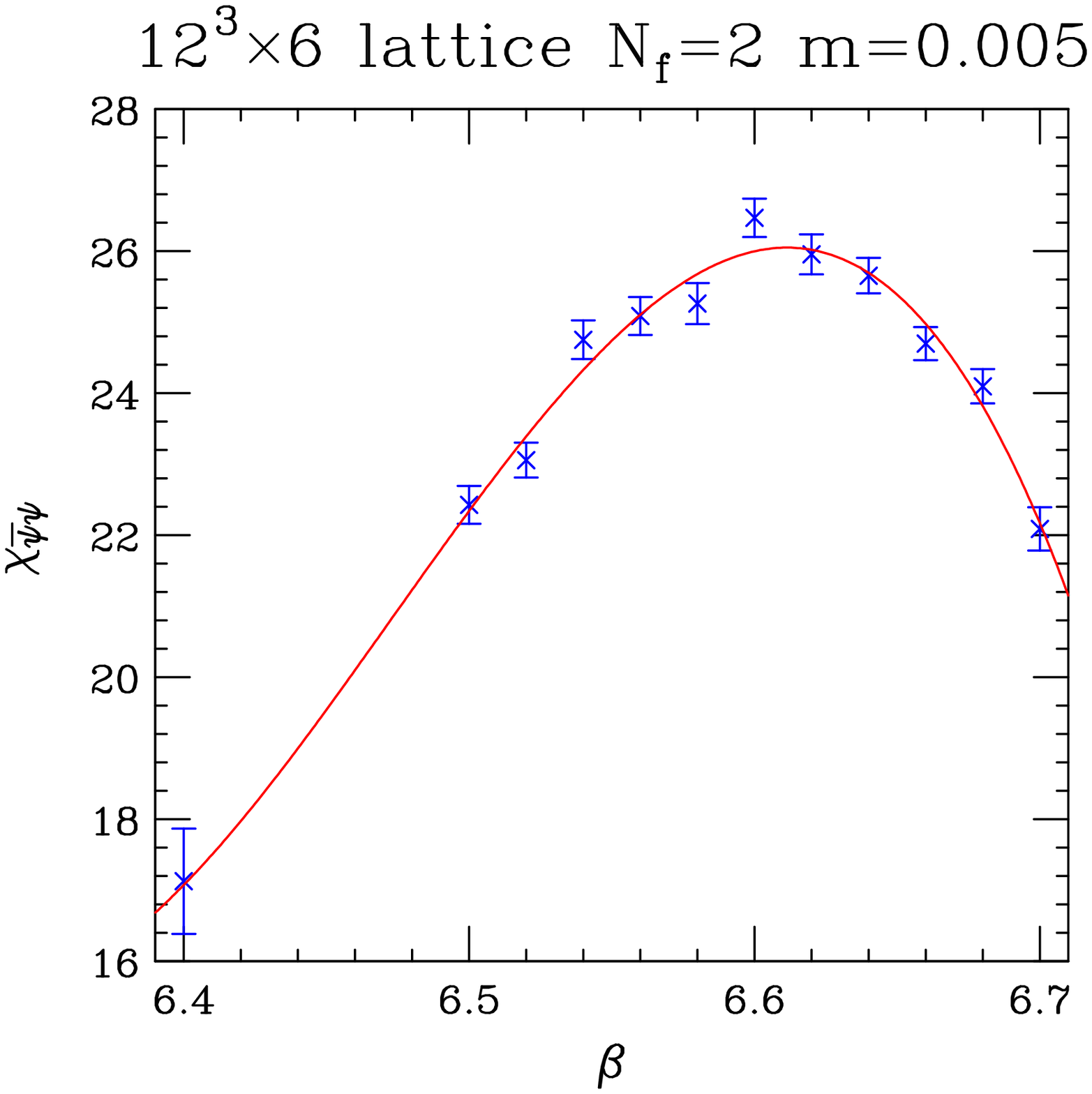}}
\caption{Chiral susceptibilities on a $12^3 \times 6$ lattice, with $N_f=2$
$m=0.005$. The curve is the fit described in the text, with $a=26.0489$, 
$b=407.578$, $c=978.166$ and $\beta_\chi=6.61143$.}
\label{fig:chi6.005}
\end{figure}

Here we attempted to determine the position of the peak of the $m=0.005$ 
susceptibility using Ferrenberg-Swendsen interpolation, but were unable to 
obtain consistent results. We therefore chose to fit the `data' with a smooth
curve:
\begin{equation}
\chi_{\bar{\psi}\psi} = a - b\,(\beta-\beta_\chi)^2 - c\,(\beta-\beta_\chi)^3 .
\label{eqn:fits}
\end{equation}
The rational for this simple form is so that we can use the same form for each
$N_t$. The second term is to give a simple parabolic fit to the peak. The third
term is necessary because, as is obvious for the larger $N_t$s, the 
susceptibility is not symmetric around the peak. The value obtained for 
$\beta_\chi$ from a fit using all $\beta$s in the range $6.4 \le \beta \le 6.7$
is $\beta_\chi=6.611(3)$ for a fit with $\chi^2/d.o.f= 1.55$, which is 
acceptable. This fit is shown, superimposed on the data in 
figure~\ref{fig:chi6.005}. 

\subsection{$N_t=8$}

\begin{figure}[htb]
\epsfxsize=6.0in
\centerline{\epsffile{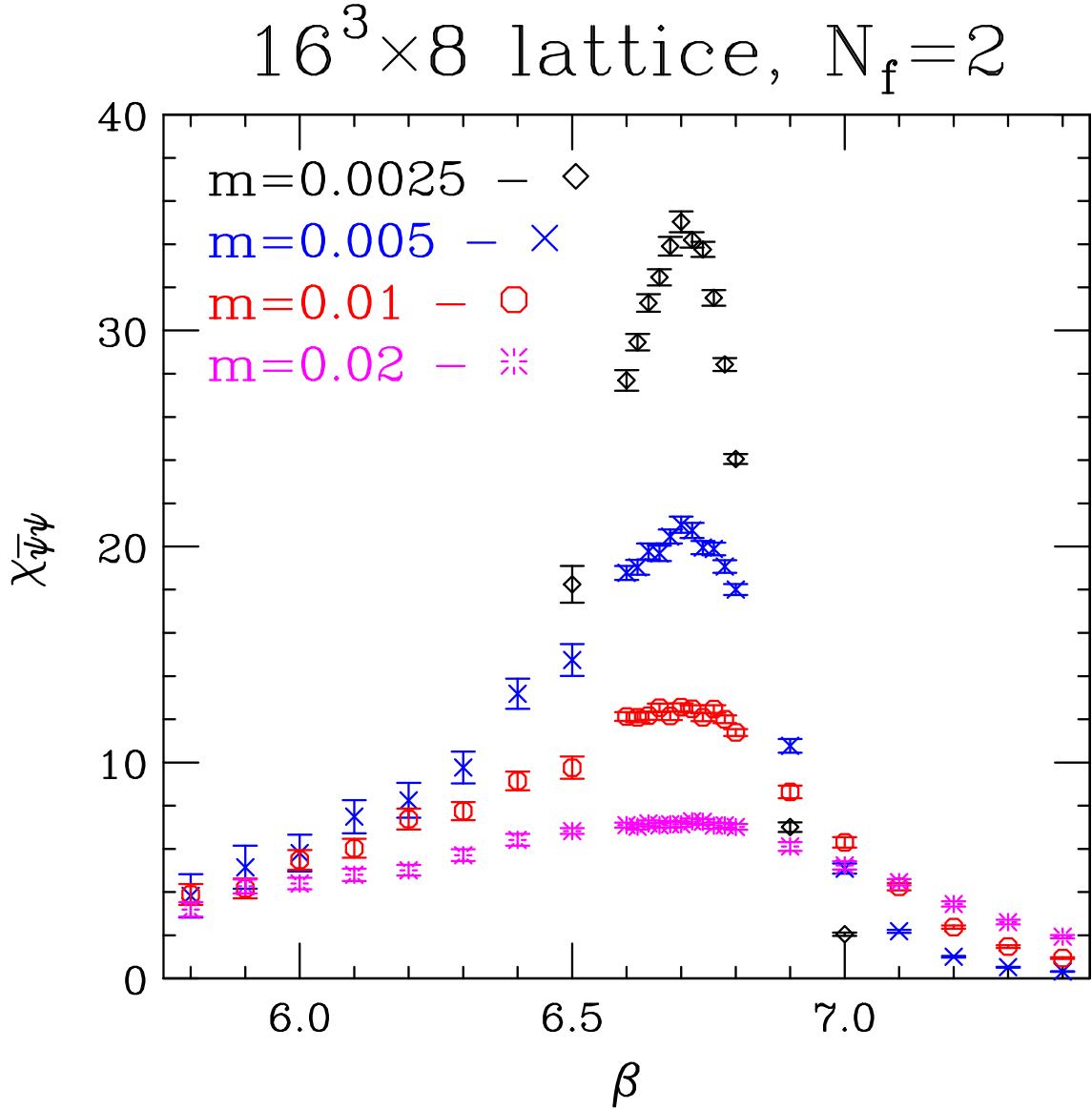}}
\caption{Chiral susceptibilities on a $16^3 \times 8$ lattice with $N_f=2$.}
\label{fig:chi8}
\end{figure}

We have extended our simulations at $N_t=8$ on a $16^3 \times 8$ lattice. For 
the 3 masses considered in our earlier work, $m=0.005$, $m=0.01$ and $m=0.02$,
we have increased the number of $\beta$ values in the neighbourhood of the
chiral transition, $6.6 \le \beta \le 6.8$, by simulating at $\beta$s separated
by $0.02$ compared with our previous $0.1$. We have increased our statistics
to 50,000 trajectories at each $(\beta,m)$ in this range. In addition, we have
simulated at a new, lower mass, $m=0.0025$. Again we have covered the range
$6.6 \le \beta \le 6.8$ with $\beta$s spaced by $0.02$, performing runs of
100,000 trajectories at each $\beta$. Outside this range we performed a run of
20,000 trajectories at $\beta=6.5$, and runs of 10,000 trajectories at 
$\beta=6.9$ and $\beta=7.0$, for this smallest mass. Figure~\ref{fig:chi8} shows
the chiral susceptibilities for these runs.

To estimate the position of the peak of the chiral susceptibility for 
$m=0.0025$, we first consider using Ferrenberg-Swendsen interpolation of the
chiral susceptibilities. This is possible, since the distributions of plaquette
values for adjacent $\beta$s in the range $6.6 \le \beta \le 6.8$ show 
significant overlap. Here we performed extrapolations from the susceptibilities
for $\beta=6.66$, $\beta=6.68$, $\beta=6.70$, $\beta=6.72$ and $\beta=6.74$,
and looked for consistency in our predictions. The best consistency we found
was between extrapolations from $\beta=6.68$, which predicted a peak at
$\beta=6.691(24)$, and $\beta=6.70$, which predicted a peak at $\beta=6.689(5)$.
Combining these we obtain a prediction $\beta_\chi=6.69(1)$.

\begin{figure}[htb]
\epsfxsize=6.0in
\centerline{\epsffile{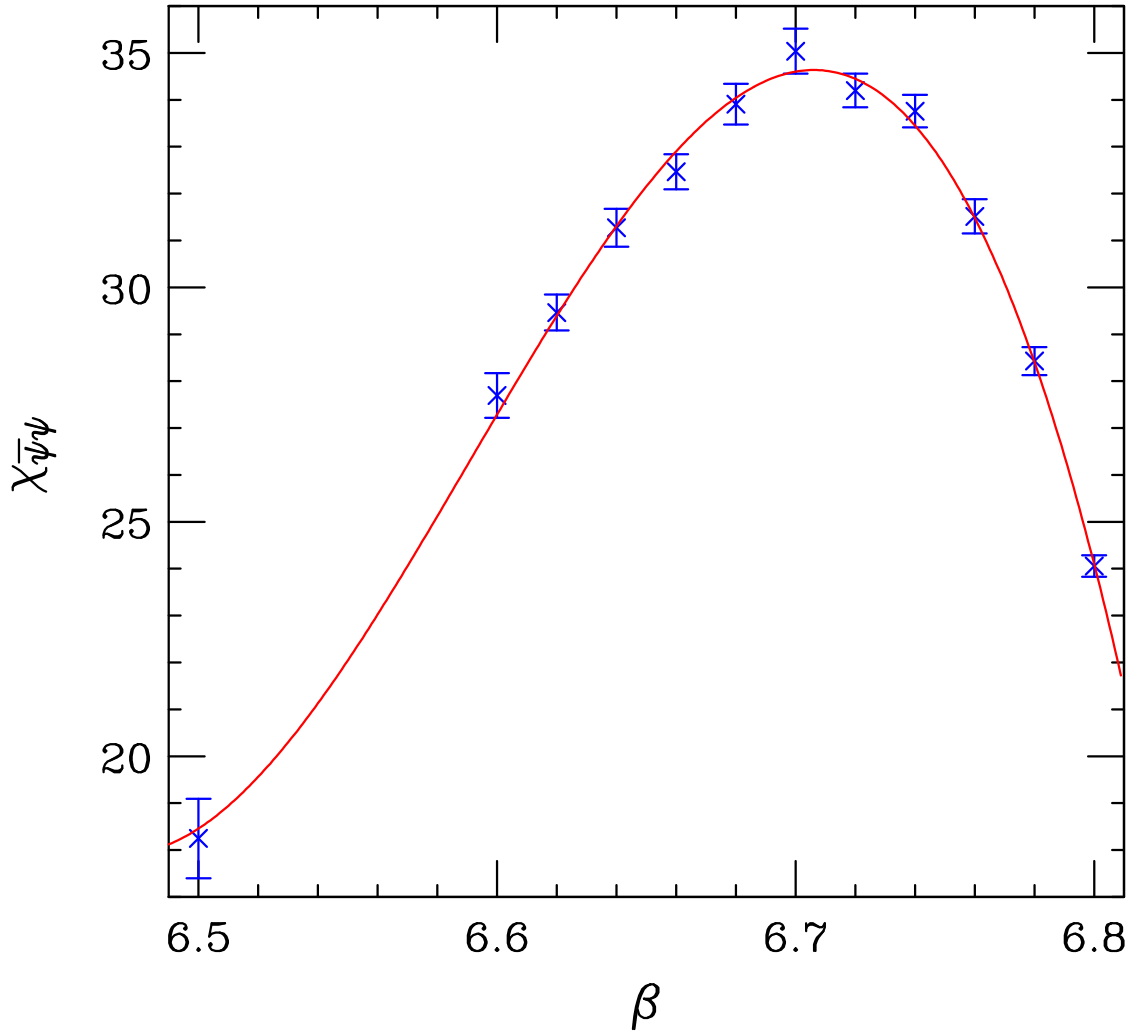}}
\caption{Chiral susceptibilities on a $16^3 \times 8$ lattice with $N_f=2$ and
$m=0.0025$. The curve is the fit described in the text, with $a=34.6359$,
$b=940.687$, $c=2716.49$ and $\beta_\chi=6.70613$.}
\label{fig:chi8.0025}
\end{figure}

A second estimate comes from fitting our susceptibilities to the form we used
for $N_t=6$ (equation~\ref{eqn:fits}). For $m=0.0025$, fitting to this cubic
polynomial over all points in the range $6.5 \le \beta \le 6.8$ yields
$\beta_\chi=6.706(1)$ for the value of $\beta$ at the peak. This fit has
$\chi^2/d.o.f=0.55$, which is excellent. Figure~\ref{fig:chi8.0025} shows this
fit superimposed on the `data'. Performing a similar fit to the susceptibilities
at $m=0.005$ over the range $6.5 \le \beta \le 6.9$ gives $\beta_\chi=6.701(4)$
with $\chi^2/d.o.f=0.85$. Fitting the susceptibilities for $m=0.01$ over the
range $6.5 \le \beta \le 6.9$ yields $\beta_\chi=6.693(8)$ with 
$\chi^2/d.o.f=1.36$ while fits to the susceptibilities for $m=0.02$ over the
same range predicts $\beta_\chi=6.71(1)$ with $\chi^2/d.o.f=0.29$. A word of
caution is due concerning the fits for the 2 largest masses. In both these 
cases, the measured susceptibility is statistically flat over an appreciable
neighbourhood of the transition. The positions of these 2 peaks is thus
determined largely by the outlying points on the fits, and should therefore not
be taken too seriously. The main reason for performing these high-mass fits is 
to get an estimate of the height of these peaks from the parameter $a$ in the 
fits.

\begin{figure}[htb]
\epsfxsize=6.0in
\centerline{\epsffile{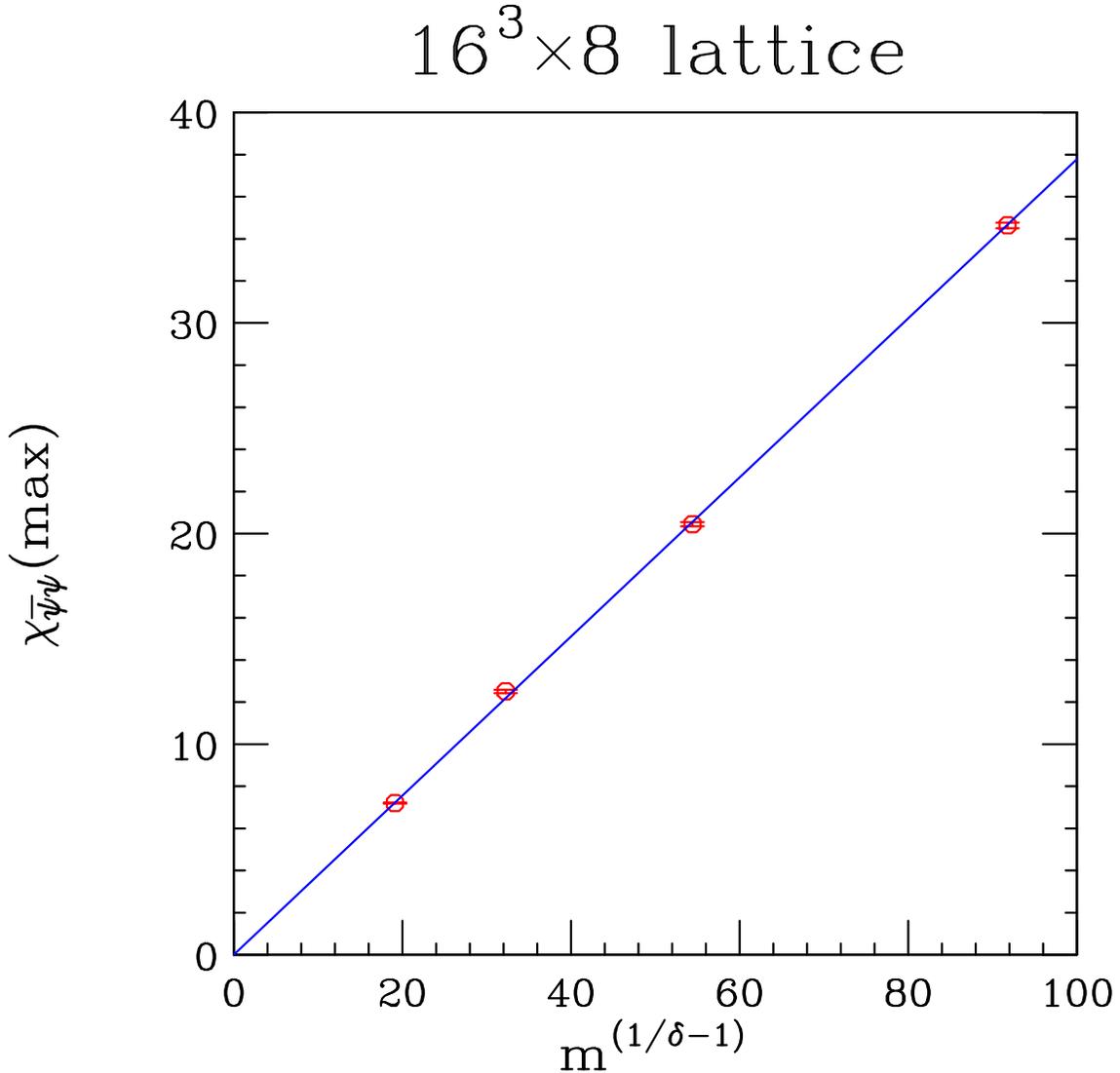}}
\caption{Peak of chiral susceptibility as a function of mass, with fit to 
critical scaling form $\chi_{max}=A m^{1/\delta-1}$ on a $16^3 \times 8$ 
lattice.}
\label{fig:chimax8}
\end{figure}

For a second-order phase transition, the value $\chi_{\bar{\psi}\psi}$ at the
peak, $\chi_{max}$ is expected to scale with mass as:
\begin{equation}
\chi_{max}=A m^{1/\delta-1} .
\label{eqn:chiscale}
\end{equation}
If the chiral transition is a finite-temperature transition, it is expected to
lie in the $O(2)$ or $O(4)$ universality class where the critical exponent
$\delta \approx 4.8$. If it is a bulk transition, which is expected to be first
order, $\delta=\infty$. The best fit to equation~\ref{eqn:chiscale} gives
$\delta=4.1(1)$ and has $\chi^2/d.o.f.=9$. Figure~\ref{fig:chimax8} shows this
fit superimposed on the values of $\chi_{max}$ taken from the values of $a$ in
the susceptibility fits. Clearly the reason that the estimated quality of the
fit (reduced $\chi^2$) is poor is because the systematic errors associated with
choosing $a$ as an estimate for $\chi_{max}$ have been ignored, whereas, 
especially for the larger masses, these clearly dominate.

At $N_t=8$, we have also performed simulations with $m=0.0025$ on a
$24^3 \times 8$ lattice at 3 $\beta$ values, near and above the chiral
transition, to check for finite volume effects. Values of the chiral
susceptibilities, which should be most susceptible to finite volume effects, are
given here, along with their values on a $16^3 \times 8$ lattice in square
brackets. For $\beta=6.7$, $\chi_{\bar{\psi}\psi}=34.4(4)\:[35.0(5)]$, for 
$\beta=6.76$, $\chi_{\bar{\psi}\psi}=30.9(4)\:[31.5(4)]$, while for $\beta=6.9$,
$\chi_{\bar{\psi}\psi}=7.0(2)\:[7.0(2)]$. These results are in good-enough
agreement for us to conclude that finite volume effects are small at the masses
we use.

\subsection{$N_t=12$}

We perform simulations on a $24^3 \times 12$ lattice at masses $m=0.0025$,
$m=0.005$ and $m=0.01$, in the neighbourhood of the chiral transition. At the
smallest mass, $m=0.0025$, we perform simulations at $\beta$ values spaced by
$0.02$ over the range $6.6 \le \beta \le 6.9$. For $6.6 \le \beta \le 6.66$ we
perform runs of 50,000 trajectories at each $\beta$. For the range
$6.68 \le \beta \le 6.9$ we perform runs of 100,000 trajectories per $\beta$. 
At $\beta=6.5$, we run for 25,000 trajectories, while for $\beta=7.0$, 
$\beta=7.1$ and $\beta=7.2$, we perform runs of 10,000 trajectories per $\beta$.
At mass $m=0.005$ we again cover the interval $6.6 \le \beta \le 6.9$ at 
increments of $0.02$. For $\beta=6.6$, we run for 100,000 trajectories. For the
rest of the interval i.e. $6.62 \le \beta \le 6.9$ we run for 50,000 
trajectories per $\beta$. At $\beta=6.5$ we run for 25,000 trajectories, while
for $\beta=6.4$, $\beta=7.0$, $\beta=7.1$ and $\beta=7.2$, we run for 10,000
trajectories per $\beta$. At $m=0.01$ we perform simulations of 25,000 
trajectories per $\beta$ at $\beta$s spaced by $0.02$ over the range
$6.6 \le \beta \le 6.9$. At $\beta=6.3,6.4,6.5$ we run for 25,000 trajectories
per $\beta$, while for $\beta=6.2$ we run for 12,500 trajectories. For 
$\beta=6.0,6.1$ and for $\beta=7.0,7.1,7.2$ we perform runs of 10,000
trajectories. We also run for 50,000 trajectories per $\beta$ for $\beta$s
spaced by $0.02$ over the range $5.7 \le \beta \le 5.9$, which is in the
neighbourhood of the deconfinement transition. These runs at $\beta < 6$ will
be discussed further in section~5.

\begin{figure}[htb]
\epsfxsize=4.0in
\centerline{{\normalsize a} \epsffile{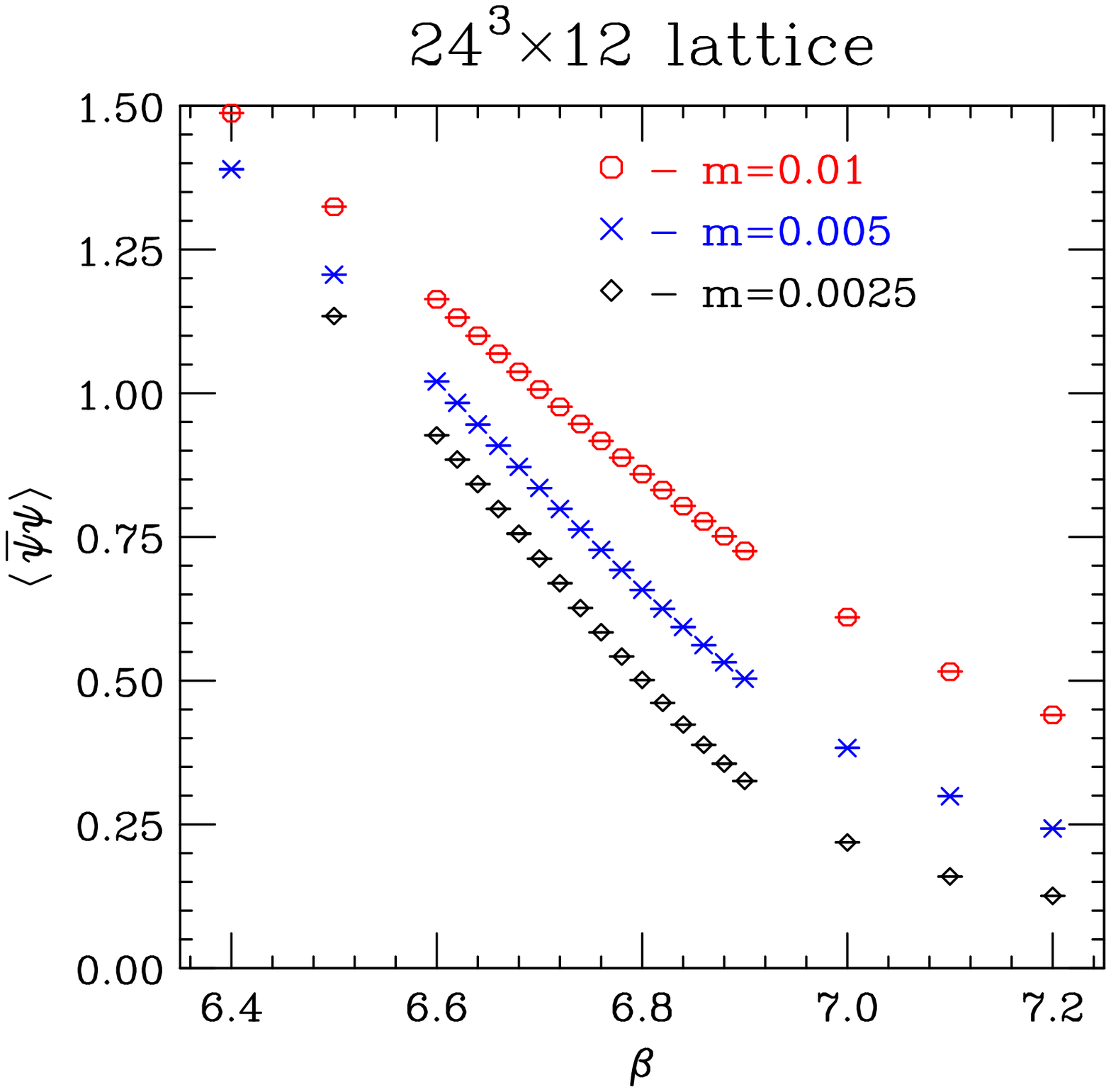}}
\epsfxsize=4.0in
\centerline{{\normalsize b} \epsffile{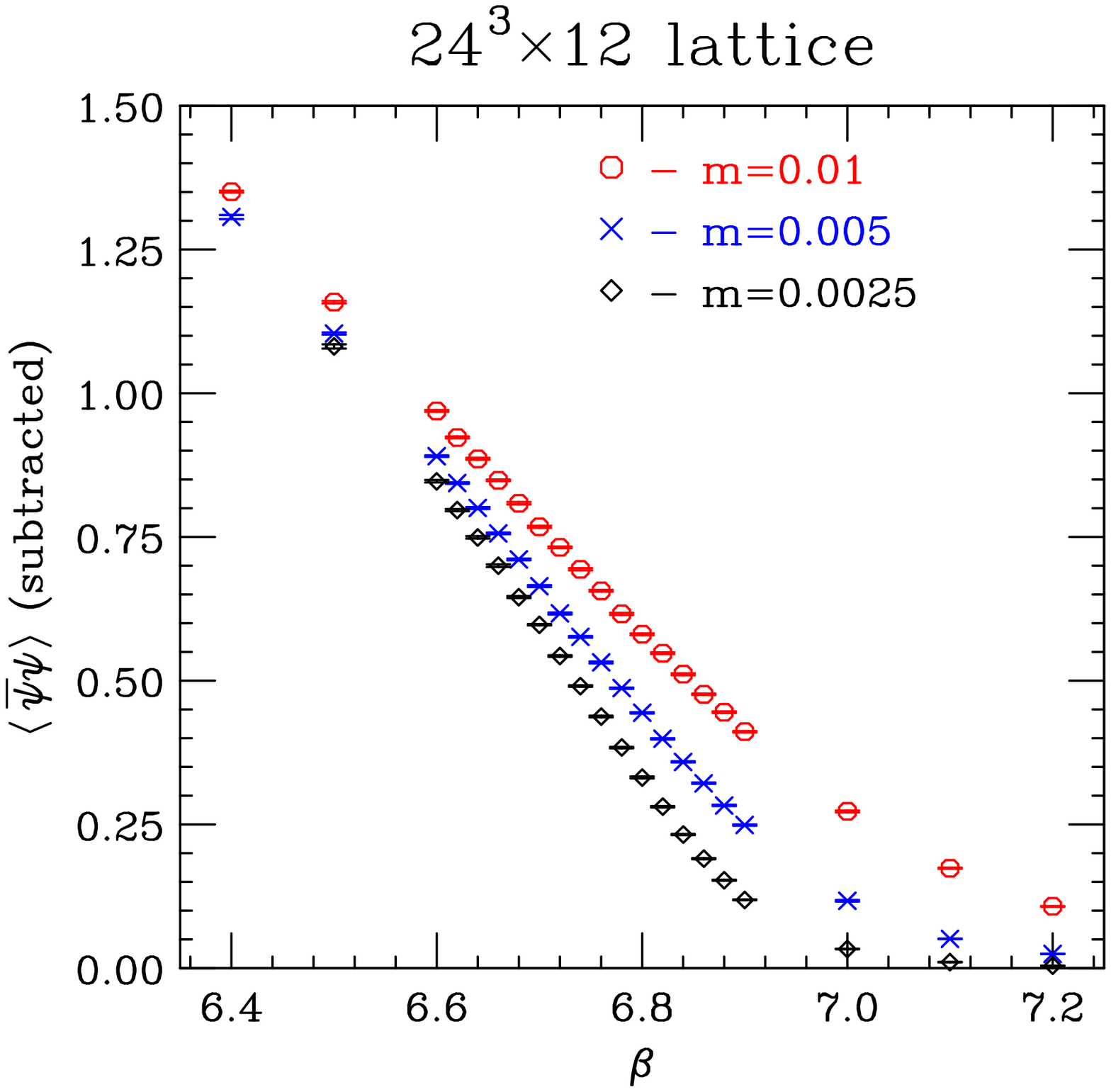}}   
\caption{a) Unsubtracted chiral condensates $\langle\bar{\psi}\psi\rangle$ as
functions of $\beta$ for masses $m=0.0025,0.005,0.01$.
b) Chiral condensates $\langle\bar{\psi}\psi\rangle$, subtracted using 
the Lattice Higgs Collaboration's prescription, as functions of $\beta$ for 
masses $m=0.0025,0.005,0.01$.}
\label{fig:pbp}
\end{figure}

Figure~\ref{fig:pbp}a shows the chiral condensates 
($\langle\bar{\psi}\psi\rangle$), as functions of $\beta$ for all 3 masses
$m=0.0025$, $m=0.005$, $m=0.01$. These are bare (lattice) quantities. However,
at non-zero mass, if we expand in powers of the quark mass $m$, the coefficient
of $m$ in physical units diverges as $1/a^2$ as $a \rightarrow 0$, and should
be regularized. We therefore subtract part of this divergence using the
prescription adopted by the Lattice Higgs Collaboration, where the subtracted
chiral condensate is defined by:
\begin{equation}
\langle{\bar{\psi}\psi}\rangle_{sub} = \langle{\bar{\psi}\psi}\rangle 
-\left(m_V\frac{\partial}{\partial m_V}\langle{\bar{\psi}\psi}\rangle\right)
                                                                  _{m_V=m}\;,
\end{equation}
where $m_V$ is the valence-quark mass. What we observe is, that while the
unsubtracted chiral condensate shows indications that it will vanish in the
chiral ($m \rightarrow 0$) limit for $\beta$ sufficiently large, the subtracted
chiral condensate shows this vanishing more clearly. However, even the 
subtracted chiral condensate does not yield an estimate for $\beta_\chi$ which
is accurate enough for our purposes. We thus turn to using the peaks of the
chiral susceptibilities, extrapolated to zero mass as our estimates for 
$\beta_\chi$.

\begin{figure}[htb]
\epsfxsize=6.0in
\centerline{\epsffile{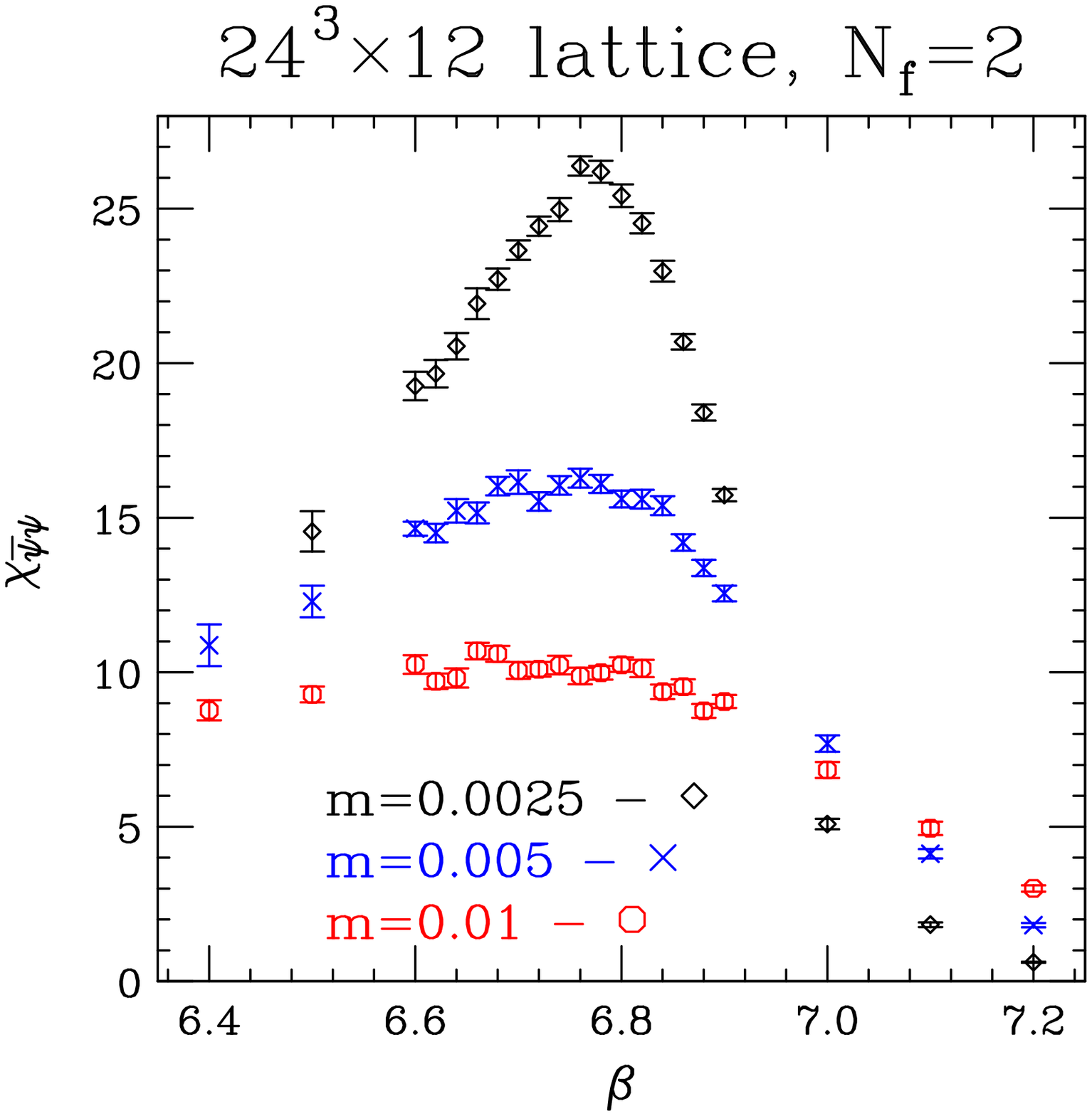}}
\caption{Chiral susceptibilities on a $24^3 \times 12$ lattice with $N_f=2$.}
\label{fig:chi12}
\end{figure}

\begin{figure}[htb]                                      
\epsfxsize=6.0in                                                         
\centerline{\epsffile{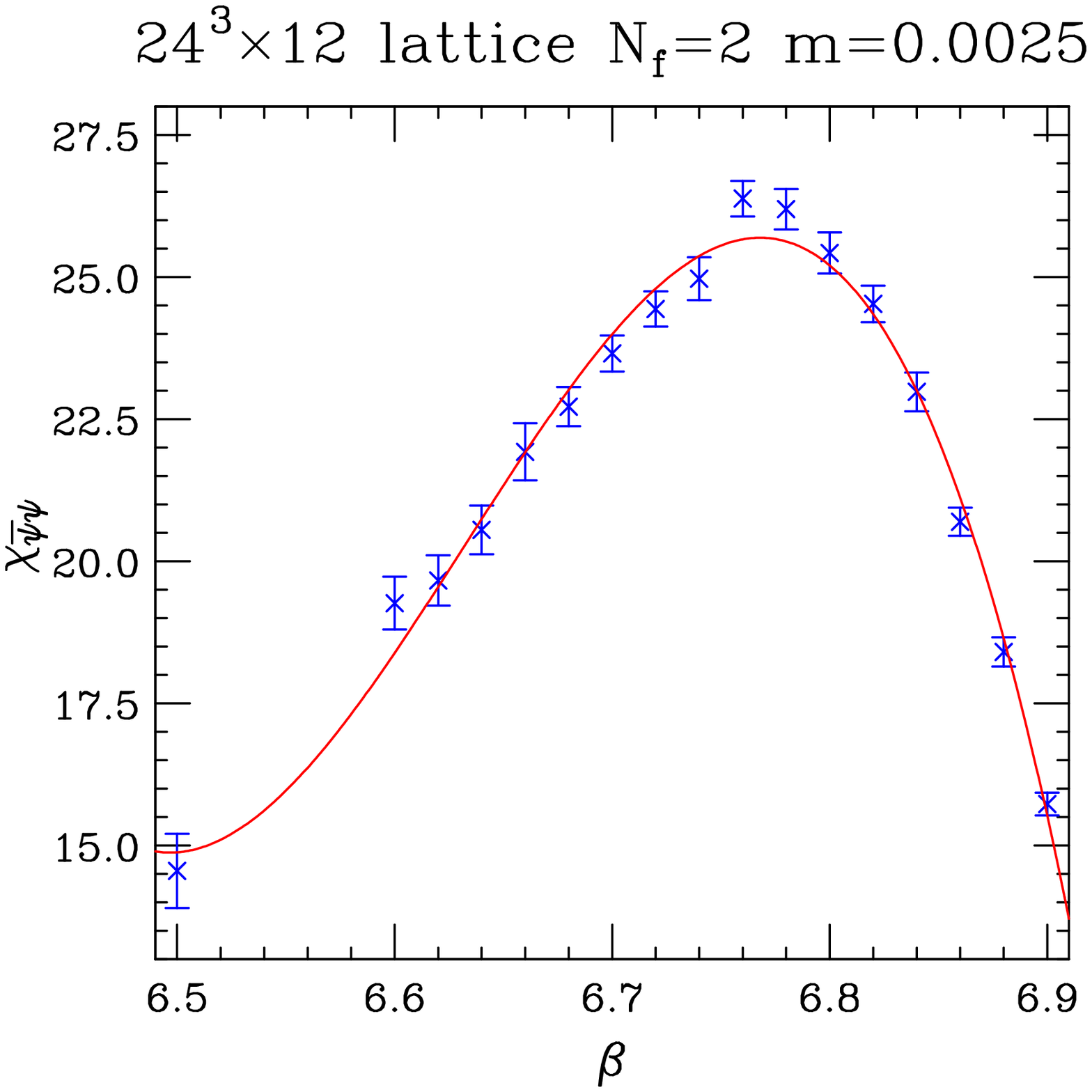}}                                    
\caption{Chiral susceptibilities on a $24^3 \times 12$ lattice with $N_f=2$ and
$m=0.0025$. The curve is the fit described in the text, with $a=25.691$,      
$b=440.589$, $c=1082.3$ and $\beta_\chi=6.76801$.}                            
\label{fig:chi12.0025}                                                      
\end{figure} 

Figure~\ref{fig:chi12} shows the chiral susceptibilities, defined in 
equation~\ref{eqn:chi} extracted from our measurements of 
$\langle{\bar{\psi}\psi}\rangle$ (5 per trajectory) in our simulations on 
$24^3 \times 12$ lattices for masses $m=0.0025$, $m=0.005$ and $m=0.01$. Here,
the distributions of plaquette values for adjacent $\beta$s have insufficient
overlap to even attempt using Ferrenberg-Swendsen interpolation to estimate the
positions of the peaks in the susceptibilities for the 3 masses. The 
susceptibilities for $m=0.0025$ show a clear peak, those for $m=0.005$ show
some indication of a rather flat peak, while those for $m=0.01$ show little
evidence for any peak. In order to extract an estimate of the positions of these
peaks, we use the fitting form used for $N_t=6$ and $8$ 
(equation~\ref{eqn:fits}), which makes maximal use of the `data'. The fit to
the $m=0.0025$ susceptibilities for all points in the range 
$6.5 \le \beta \le 6.9$ yields $\beta_\chi=6.768(2)$ in an acceptable fit with 
$\chi^2/d.o.f.=1.68$. This fit is superimposed over the measured 
susceptibilities in figure~\ref{fig:chi12.0025}. A fit to the $m=0.005$ `data'
over the same interval predicts $\beta_\chi=6.745(6)$ with $\chi^2/d.o.f.=0.91$,
while a fit to the $m=0.01$ data also over the same range gives 
$\beta_\chi=6.70(3)$ with $\chi^2/d.o.f.=1.52$. These last 2 fits should not
be considered too seriously, because these peaks are defined by outlying rather
than central points, owing to the flatness of the distributions. Their main
purpose is to yield an estimate for the values of the parameter $a$ at their
peaks.

\begin{figure}[htb]                                                             
\epsfxsize=6.0in                                                                
\centerline{\epsffile{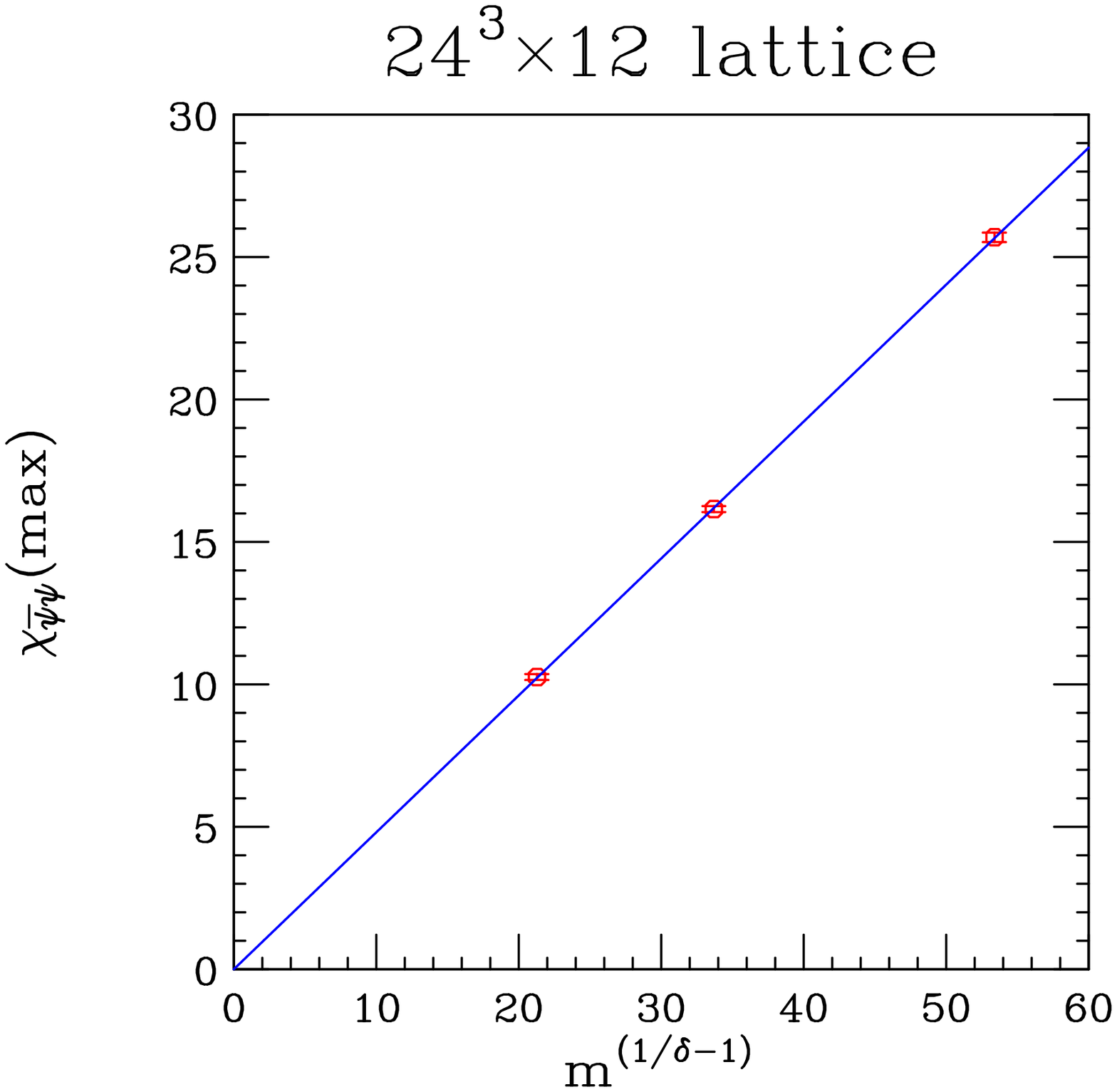}}                                   
\caption{Peak of chiral susceptibility as a function of mass, with fit to      
critical scaling form $\chi_{max}=A m^{1/\delta-1}$ on a $24^3 \times 12$
lattice.}                          
\label{fig:chimax12}                                                            
\end{figure}     

We also consider the scaling properties of the susceptibility peak with mass
using the scaling form of equation~\ref{eqn:chiscale}. With only 3 points, there
is only one degree of freedom. Our fit gives the critical exponent 
$\delta=2.98(4)$ with $\chi^2/d.o.f.=0.29$. This is in agreement with the
mean-field (free-field) critical exponent $\delta=3$, rather than the expected
$O(2)$ or $O(4)$ critical exponent $\delta \approx 4.8$ or first-order scaling
$\delta=\infty$. The graph of our data -- the values of $a$ from our fits --
with this fit superimposed is given in figure~\ref{fig:chimax12}.

\section{Comparison with perturbative predictions}

Here for consistency we only consider the values of $\beta_\chi$ obtained from
fitting our chiral susceptibilities to the form given in 
equation~\ref{eqn:fits}. The errors given in the previous section do not 
include any systematic errors due to this rather arbitrary choice of a fitting
function or to the selection of the range of $\beta$ values over which the
fits are performed. We will assume $0.01$ as a conservative estimate of these
systematic errors, at least for the lightest masses. Because our measurements
are inadequate to reliably determine if there is any significant mass 
dependence on the position of the peaks, we take our measurement of the position
of the peak for the lightest mass as our estimate for the position of the
transition in the chiral limit. We assume that our estimate of the systematic
errors is large enough to encompass the shift in the positions of the peaks as
we approach the chiral limit. Table~\ref{tab:trans} summarises the results of 
our determinations of the positions of the chiral transitions ($\beta_\chi$)
from this and previous publications, as well as those of the positions of the
deconfinement transitions ($\beta_d$) from previous publications and the next
section.

\begin{table}[h]
\centerline{
\begin{tabular}{|c|l|l|}
\hline
$N_t$   & \multicolumn{1}{c|}{$\beta_d$} & \multicolumn{1}{c|}{$\beta_\chi$} \\
\hline
4              &$\;$5.40(1)$\;$  &$\;$6.3(1)$\;$            \\
6              &$\;$5.54(1)$\;$  &$\;$6.61(1)$\;$           \\
8              &$\;$5.65(1)$\;$  &$\;$6.71(1)$\;$           \\
12             &$\;$5.81(1)$\;$  &$\;$6.77(1)$\;$           \\
\hline
\end{tabular}
}
\caption{$N_f=2$ deconfinement and chiral transitions for $N_t=4,6,8,12$.}
\label{tab:trans}
\end{table}

Equation~\ref{eqn:deltabeta} gives the perturbative asymptotic freedom
predictions for the changes in $\beta_\chi$ between different $N_t$s. In 
particular:
\begin{eqnarray}
\beta_\chi(N_t=8 )-\beta_\chi(N_t=6) &\approx& 0.087 \nonumber \\
\beta_\chi(N_t=12)-\beta_\chi(N_t=8) &\approx& 0.122           \\
\beta_\chi(N_t=12)-\beta_\chi(N_t=6) &\approx& 0.209 \nonumber \:.
\end{eqnarray}
The reason for including the third equation is because, if this is a finite
temperature transition, the fermion mass in physical units for $m=0.005$ at
$N_t=6$ and for $m=0.0025$ at $N_t=12$ are identical, so that one might hope
that any error due to not extrapolating to the chiral limit might be minimized.
Our measurements (from table~\ref{tab:trans}) give
\begin{eqnarray}                                                          
\beta_\chi(N_t=8 )-\beta_\chi(N_t=6) &=& 0.10(1) \nonumber \\          
\beta_\chi(N_t=12)-\beta_\chi(N_t=8) &=& 0.06(1)           \\          
\beta_\chi(N_t=12)-\beta_\chi(N_t=6) &=& 0.16(1) \nonumber \:.          
\end{eqnarray}
If taken at face value, these favour the conformal option, where the fact that
$\beta_\chi(N_t=12)-\beta_\chi(N_t=8)$ is roughly half the value predicted by
asymptotic freedom could indicate that $\beta_\chi$ is approaching a non-zero
limit.

We need to be cautious, since the lattice coupling and hence $1/\beta$ is known
to be a poor expansion parameter. That is, higher order terms for the expansion
of any quantity in powers of $g^2$ tend to be large. The simplest improved
choice of $\beta$s is the tadpole-improved $\beta$ of Lepage and Mackenzie
\cite{Lepage:1992xa}:
\begin{equation}
\bar{\beta} = \frac{1}{3}\langle{\rm Tr}_\Box UUUU\rangle\beta \:.
\end{equation} 
For connection with Lepage-Mackenzie,
$\bar{\beta}=6/\bar{g}^2=6/4\pi\bar{\alpha}$. (Note that for staggered
fermions, tadpole improvement of the fermion determinant is equivalent to
rescaling the fermion mass and can thus be ignored, since we are interested in
the limit $m \rightarrow 0$.) The plaquette in the above equation should be
evaluated at $\beta$, on a lattice which is at zero temperature. Since $T=0$,
in practice means on an $N_t^4$ lattice for which $\beta << \beta_d(N_t)$, this
would require simulating on lattices much larger than any we contemplate. For
this reason, we use the finite temperature plaquettes from our simulations in
this equation. This yields $\bar{\beta}_\chi(N_t=6)=4.48(1)$,
$\bar{\beta}_\chi(N_t=8)=4.58(1)$ and $\bar{\beta}_\chi(N_t=12)=4.65(1)$. This
gives
\begin{eqnarray}
\bar{\beta}_\chi(N_t=8 )-\bar{\beta}_\chi(N_t=6) &=& 0.10(1) \nonumber\\
\bar{\beta}_\chi(N_t=12)-\bar{\beta}_\chi(N_t=8) &=& 0.07(1)          \\
\bar{\beta}_\chi(N_t=12)-\bar{\beta}_\chi(N_t=6) &=& 0.17(1) \nonumber\:.
\end{eqnarray} 
compared with the perturbative prediction:
\begin{eqnarray}
\bar{\beta}_\chi(N_t=8 )-\bar{\beta}_\chi(N_t=6) &\approx& 0.083 \nonumber\\
\bar{\beta}_\chi(N_t=12)-\bar{\beta}_\chi(N_t=8) &\approx& 0.117          \\
\bar{\beta}_\chi(N_t=12)-\bar{\beta}_\chi(N_t=6) &\approx& 0.200 \nonumber\:.
\end{eqnarray} 
While this is an overall improvement, it is insufficient. Choosing instead
$\beta_V$, the $\beta$ associated with the inter-quark potential, makes little
difference. Here $\beta_V=6/g_V^2=6/4\pi\alpha_V$. We use the relation between
$\bar{\alpha}$ and $\alpha_V$ from Lepage-Mackenzie to obtain $\beta_V$. The
reason that going from $\beta$ to $\beta_V$ does not make much difference is
because, as noted by Lepage-Mackenzie, the perturbative relation:
\begin{equation}
\beta=\beta_V+2.245+{\cal O}(1/\beta_V)
\end{equation}
(where we have chosen the momentum scale at which we measure $\beta_V$ to be
$\pi/a$), is a good approximation. With such a constant shift, differences in
$\beta$s are left unchanged. In the prediction, based on the perturbative
$\beta$-function, for the $\beta$s ($\beta_V$s) we consider, the 2-loop
contribution is small, so the perturbative predictions for differences in
$\beta$s and $\beta_V$s are almost the same. Here we see that replacing the
lattice coupling with an improved coupling such as $g_V$ does not significantly
affect changes in $\beta$, since $g$ is small enough that the 2-loop 
contribution to the $\beta$-function is significantly smaller then the 1-loop
contribution, for both the original and improved schemes. Remember that the
1- and 2-loop coefficients in the $\beta$-function are scheme independent. 
Hence changing from lattice to improved couplings will not significantly
improve agreement between measured and predicted running of the couplings.
Of course, choosing a {\it much} smaller momentum scale for $\beta_V$, driving 
it towards the perturbative fixed point, could improve agreement for
$\beta_\chi(N_t=12)-\beta_\chi(N_t=8)$, but would be difficult to justify.

However, it is well-known that even with tadpole improvement of the gauge links,
perturbation theory for staggered fermions is still badly behaved 
\cite{Patel:1992vu,Golterman:1998jj}. The reason is another form of tadpole,
the `doubler tadpole' due to flavour(`taste')-mixing responsible for taste
breaking. Unfortunately for us, before an improved perturbation theory could be
developed for staggered fermions, interest shifted to improved staggered
fermions designed to minimize taste breaking, making perturbation theory
better behaved.

It is interesting to note that the lack of significant improvement using
$\bar{\beta}$ (or $\beta_V$ or $\beta_{\overline{MS}}$) instead of $\beta$ has
also been noticed by \cite{Deuzeman:2008sc} in their studies of QCD with 8
fundamental quarks using an improved staggered-quark action, so perhaps it is 
a property of theories with slowly varying running couplings and not an
artifact of using unimproved actions. This contrasts with the one system where
there are precise measurements of the finite temperature transition for a
large range of $N_t$ values, namely pure $SU(3)$ Yang-Mills theory (quenched
QCD) transcribed to the lattice using the Wilson (plaquette) action. For this
system using an improved coupling greatly improves the agreement between the
measured $N_t$ dependence of the critical coupling $\beta_c$ and the prediction
from the 2-loop $\beta$-function, as shown in table~\ref{tab:quenched}.
\newline 
\begin{table}[h]                                                              
\centerline{                                                                
\begin{tabular}{|cc|c|c|}                                                      
\hline                  
$N_t$ & $N_t'$ & $1-\frac{[\beta_c(N_t')-\beta_c(N_t)]_{\it lattice}}
{[\beta_c(N_t')-\beta_c(N_t)]_{\it 2-loop}}$ &
$1-\frac{[\bar{\beta}_c(N_t')-\bar{\beta}_c(N_t)]_{\it lattice}} 
{[\bar{\beta}_c(N_t')-\bar{\beta}_c(N_t)]_{\it 2-loop}}$                \\
\hline
  6 &  8 & 34\% & 19\%                                                   \\
  8 & 12 & 24\% & 13\%                                                   \\
 12 & 18 & 17\% &  8\%                                                   \\
\hline
\end{tabular}
}
\caption{Difference between the changes in the lattice $\beta_c$ with change
in $N_t$ and the prediction from the 2-loop perturbative $\beta$-function, 
compared with the same quantity using the improved $\bar{\beta}_c$, for 
quenched lattice QCD.}
\label{tab:quenched}
\end{table}
For these calculations we used the values of $\beta_c(N_t)$ for quenched 
lattice QCD given in the recent publication~\cite{Francis:2015lha}.

\section{The $N_t=12$ deconfinement transition.}

In section~3 we mentioned that we have extended our simulations on a 
$24^3 \times 12$ lattice at $m=0.01$ into the neighbourhood of the deconfinement
transition. Although $\beta_d$ is too small for its evolution to be governed
by perturbation theory, knowledge of its value as a function of $N_t$ is
necessary when choosing $\beta$ values for zero temperature simulations. It is
also useful to know the value of the deconfinement temperature as well as the
chiral-symmetry restoration temperature in physical units i.e. in terms of
$f_\pi \approx 246$~GeV. We chose to simulate this regime at only one quark
mass so that we could devote most of our resources to the proximity of the
chiral transition. For this same reason we chose the highest of our 3 masses.
Here we are relying on the observation based on our studies at smaller $N_t$s
that $\beta_d$ depends only weakly on the quark mass.

\begin{figure}[htb]
\epsfxsize=6.0in
\centerline{\epsffile{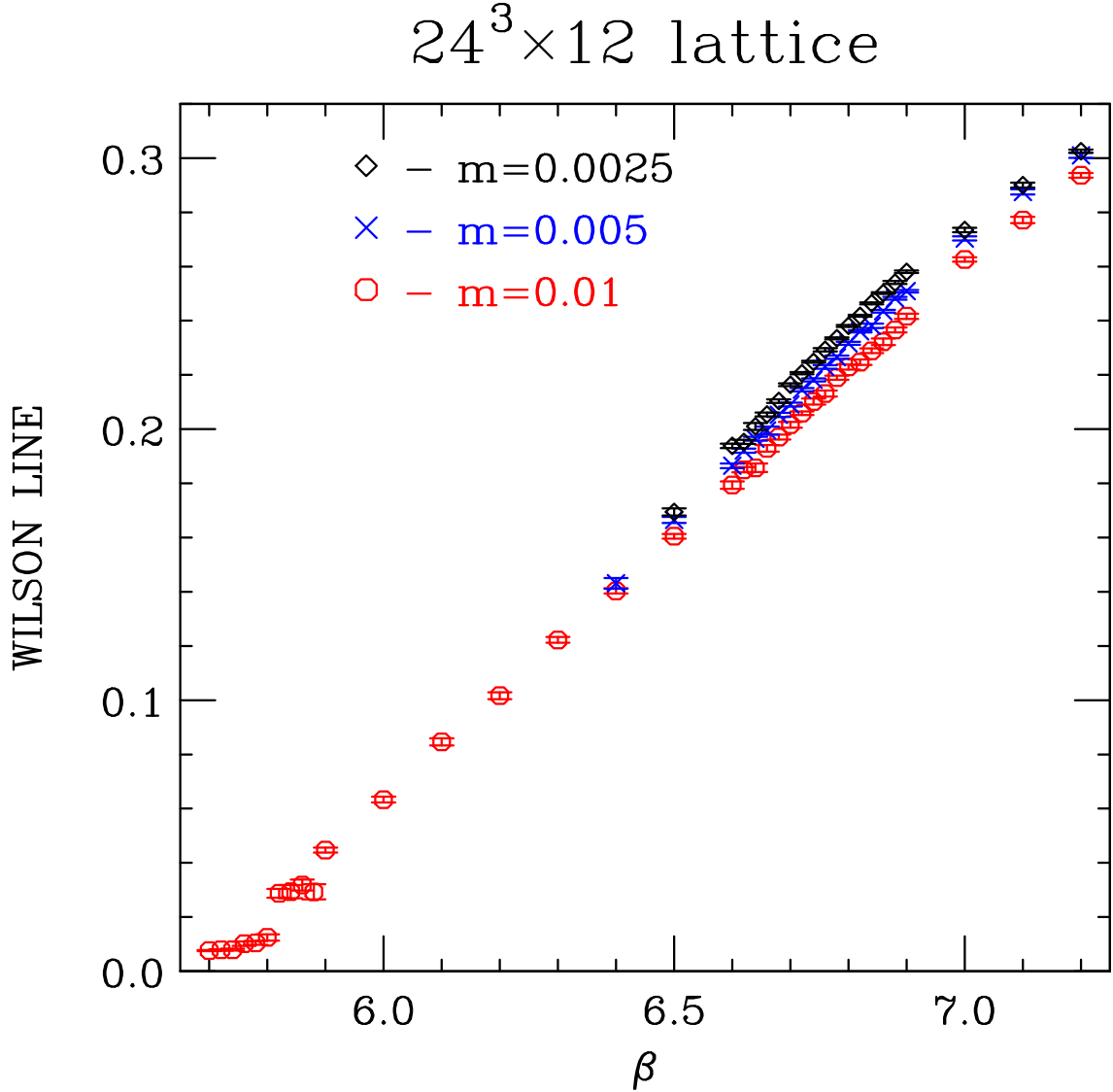}}
\caption{Wilson Lines as functions of $\beta$ on a $24^3 \times 12$ lattice.
These are traces of products gauge links in the colour-triplet representation 
of $SU(3)_{colour}$.}
\label{fig:wilson}
\end{figure}

The position of the deconfinement transition is determined by the point below 
which the Wilson Line (Polyakov Loop) becomes very small. 
Figure~\ref{fig:wilson} is a plot of Wilson Lines against $\beta$ for our
$24^3 \times 12$ simulations at all 3 masses. For the $m=0.01$ plot, we notice
that the Wilson Line is near zero for small $\beta$s and then jumps to a value
appreciably greater than zero at $\beta \approx 5.8$.

\begin{figure}[htb]                                                           
\epsfxsize=6.0in                                                             
\centerline{\epsffile{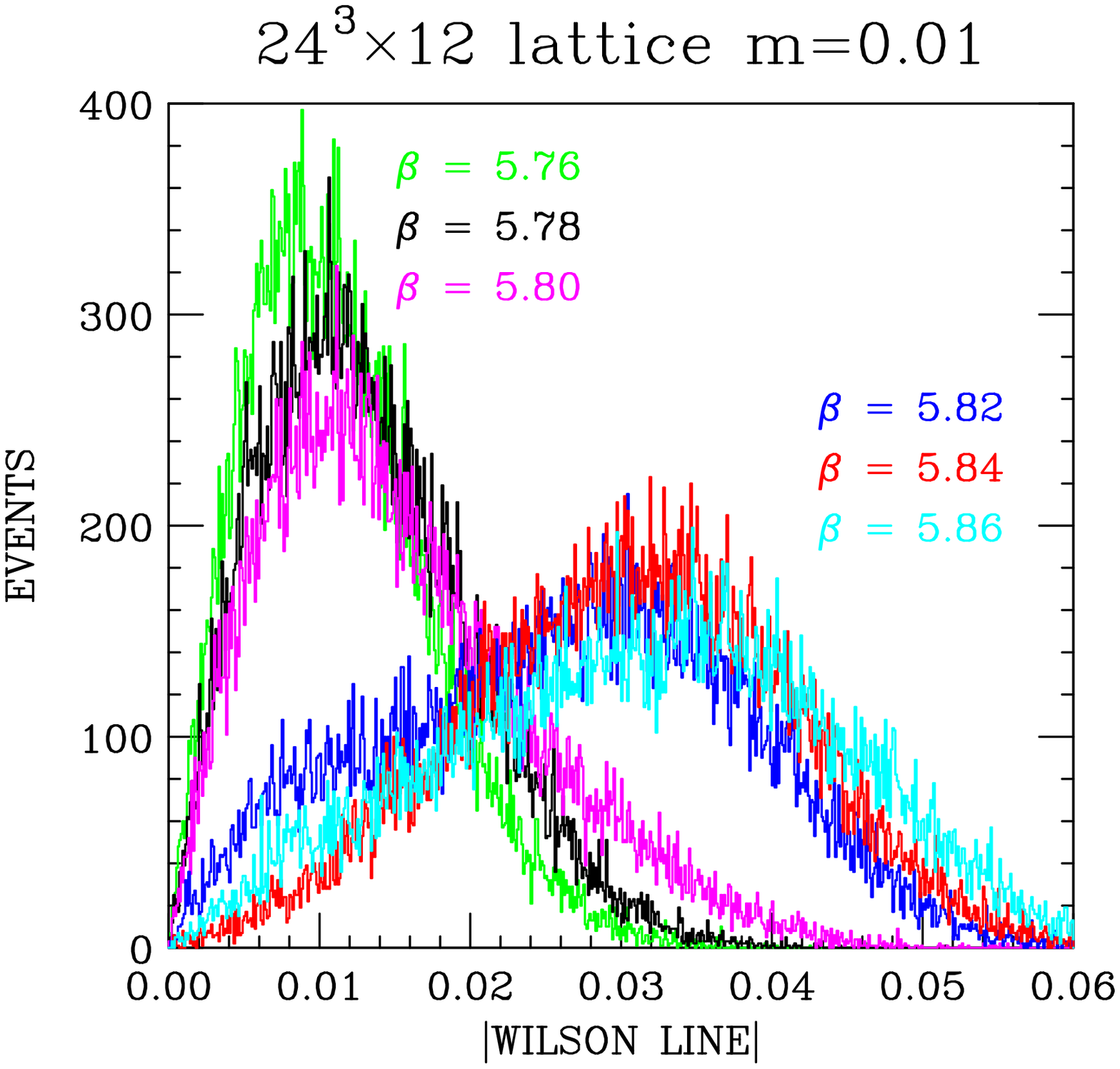}}             
\caption{Histograms of magnitudes of the Wilson Line for $\beta$ values close
to the deconfinement transition on a $24^3 \times 12$ lattice with $m=0.01$.}
\label{fig:wilsonhist}
\end{figure}

To determine the position of this deconfinement transition more precisely,
we ran for 50,000 trajectories per $\beta$ for $\beta$ values spaced by $0.02$
over the interval $5.7 \le \beta \le 5.9$. Figure~\ref{fig:wilsonhist} shows
histograms of the distribution of magnitudes of the Wilson Line, for $\beta$
values close to the deconfinement transition. 
Note the qualitative difference between these histograms for $\beta \le 5.80$
and those for $\beta \ge 5.82$. Those histograms for the lower range of $\beta$s
are peaked at $\approx 0.01$, while those for $\beta$s in the upper range are
peaked at $\approx 0.03$. From this we estimate that the deconfinement 
transition occurs at $\beta=\beta_d=5.81(1)$. The transition is very abrupt,
suggestive of a first-order phase transition. This value for $\beta_d$ has been
entered in table~\ref{tab:trans} in section~4.

\section{Discussion and conclusions}

We use studies of scaling of the assumed finite-temperature chiral transition
of lattice QCD with 2 colour-sextet quarks to attempt to determine whether
this theory is conformal or walking. This provides an alternative to the use
of step-scaling methods to achieve this goal. Recent extensive step-scaling
studies using improved staggered quarks \cite{Fodor:2015zna} indicate that
this theory walks, while similar studies using improved Wilson quarks 
\cite{Hasenfratz:2015ssa} present evidence for a fixed point, which would mean
that the theory is conformal. Hence use of a different method to try and
determine which is the correct behaviour is warranted.

We simulate lattice QCD with 2 flavours of colour-sextet quarks on lattices
with $N_t=6$, $8$ and $12$ in the vicinity of the chiral-symmetry restoration
transition, to accurately determine the value $\beta_\chi$ of $\beta=6/g^2$
at that transition. Under the assumption that this is a finite-temperature
phase transition, which thus occurs at a fixed temperature in physical units,
this gives the running of the gauge coupling $g$ as the lattice spacing $a$ is 
decreased. We compare the evolution of the coupling $g_\chi$ at this transition
with the prediction from the 2-loop perturbative $\beta$-function. What we find
is that the change in $\beta_\chi=6/g_\chi^2$ is in approximate agreement with
the perturbative prediction between $N_t=6$ and $8$, but is smaller by about a
factor of 2 than the prediction between $N_t=8$ and $12$. This suggests that
this chiral transition is a bulk transition for which $\beta_\chi$ would 
approach a finite value in the large $N_t$ limit. In this case this theory
would be a conformal field theory, and not the desired walking theory. On the
other hand, the scaling of the susceptibility peaks with mass suggests that
the chiral transition is second order, whereas the simplest scenario for a bulk
transition suggests that it should be first order. However, the fact that the
critical exponent $\delta \approx 3$ suggests that the transition is mean-field,
which would be more likely for a true 4-dimensional and hence bulk transition,
than for a finite-temperature and hence quasi-3-dimensional transition, which
should be in the universality class of the 3-dimensional $O(2)$ or $O(4)$ spin
model with $\delta \approx 4.8$.

In our initial comparison, we used the prediction in terms of the bare lattice
coupling $g(a)$, which is known to be a poor expansion parameter. We therefore
looked at the tadpole-improved coupling $\bar{g}$, which is supposed to be a 
better expansion parameter, as well as the related coupling $g_V$ which is
related to the heavy-quark potential. Both these improvements make a slight
improvement in the running of the coupling, but not nearly enough to produce
agreement with the perturbative results. (Based on our experience with quenched
QCD we would have expected an agreement to within $20\%$ or probably better.)
We do note, however, that even when such improvements are applied,
perturbation theory with unimproved staggered quarks does not work well. In
addition, the examples where use of such couplings ($\bar{g}$, $g_V$ or
$g_{\overline{MS}}$) has improved the behaviour of lattice perturbation theory
have been for theories such as QCD with quarks in the fundamental
representation of the colour group, where the massless theory has only one
mass or length scale, the scale associated with confinement and chiral
symmetry breaking. For QCD with colour-sextet quarks, we have shown that the
scales of confinement and chiral-symmetry breaking are very different. It is
unclear if methods which work for a single-scale theory will continue to work
for a two-scale theory.

We note that use of staggered quarks and, in particular, unimproved staggered
quarks, has potential difficulties because flavour symmetry (sometimes referred 
to as taste symmetry) is explicitly broken, and is only restored in the 
continuum limit. This means that in the chiral limit at non-zero lattice 
spacing, there are less massless degrees of freedom than in the continuum limit.
Because of this, it is possible that a theory with an infrared fixed point and
hence conformal, could appear to be walking. Hence, care needs to be taken in
taking the continuum and chiral limits. Using an improved action can reduce 
these problems. However, improved staggered fermions have their own 
difficulties. Improving the staggered-fermion action can introduce extra
phases which are lattice artifacts. This was observed in studies of the
apparent bulk transition in QCD with 12 fundamental quarks  
\cite{Cheng:2011ic,Deuzeman:2012ee} and provides extra complications for
studying the finite temperature behaviour of QCD with many fundamental quarks
\cite{Deuzeman:2008sc,Deuzeman:2009mh,Schaich:2012fr,Schaich:2015psa}. 
Improvement is designed to produce actions whose weak coupling physics is
closer to that of the continuum theory. However, it does not necessarily produce
better behaviour in the intermediate or strong-coupling domain.

Another word of caution is necessary concerning our results. We have only used
one spatial size ($24^3$) for our $N_t=12$ simulations, so we have not ruled out
finite volume effects. The mass dependence of the position of the peaks in
chiral susceptibilities has not been adequately explored to check that we are
really seeing the chiral limit. We have only used one action, and have not
explored whether improved actions such as those used by the Lattice Higgs
Collaboration and Degrand, Shamir and Svetitsky might show different
behaviour.

While the most direct way of answering these questions would be to continue our
work to lattices with larger $N_t$ values and smaller quark masses, such 
simulations would be expensive, and it is not clear if repeating our studies
with $N_t=16$ or $18$ would provide the desired clarification. There are,
however, other less expensive studies which could potentially help clarify the
situation. The first, which we are pursuing, is to extend our simulations of QCD
with 3 colour-sextet quarks to $N_t=12$. This theory is almost certainly 
conformal, and even if it is not, the perturbative evolution of its coupling
is extremely slow (this is because asymptotic freedom is lost at 
$N_f=3\frac{3}{10}$ for QCD with $N_f$ sextet-quark flavours. Hence we should
expect to see essentially no change in the value of $\beta_\chi$ between 
$N_t=8$ and $N_t=12$. If this is observed it would indicate that the $N_f=2$
and $N_f=3$ theories are behaving rather differently as would be expected if
the $N_f=2$ theory walks.

One method of testing how well some of the different choices of improved 
couplings work with sextet quarks and its 2 length scales would be to perform
an extensive study of the position of the chiral transition with sextet quarks
in the quenched theory. That this transition is separated from the deconfinement
transition was shown in very early studies \cite{Kogut:1984sb}. The advantage
of this approach is that the production of very large quenched lattices can be
performed very cheaply, and multimass inversions, already used in the RHMC
algorithm will allow us to study the chiral condensate over a large range of
masses. Here it will be possible compare the evolution of $\beta_d$ and
$\beta_\chi$ with various improved couplings.

Another direction we are pursuing (we thank Julius Kuti for this suggestion) is
to simulate QCD with 2 colour-sextet quarks at a fixed $\beta$ value above 
$\beta_\chi(N_t=12)$ -- we choose $\beta=6.9$, -- and simulate on lattices
with a fixed spatial volume, varying $N_t$ to look for the transition. 
Perturbation theory predicts that $\beta_\chi=6.9$ for some $N_t$ in the range
$18 < N_t < 20$. We are simulation on $24^3 \times N_t$ lattices with $8 \le
N_t \le 24$ with masses as low as $m=0.00125$. Indications are that we will
need to increase the spatial box size to accommodate the larger $N_t$s, since
we are seeing finite volume effects at $N_t=22$ and $24$.

Temporarily ignoring questions as to whether this theory is QCD-like, we have
started zero-temperature simulations at $\beta=5.81$ ($\beta_d$ for $N_t=12$),
on $24^3 \times 48$ lattices. One of the reasons for this parameter choice, is
to answer another question posed by Julius Kuti, who asked what value we
estimate for $T_d/f_\pi$. So far we have produced 250 lattices separated by
100 trajectories for $m=0.01$ ($m_\pi \approx 0.25$) and 250 lattices for 
$m=0.005$ ($m_\pi \approx 0.175$). Larger-lattice zero-temperature simulations 
at weaker couplings are being contemplated.

Note that all the simulations reported in this paper are in the state where the
argument of the Wilson Line (Polyakov Loop) is close to zero. Those states 
where the argument of the Wilson Line is close to $\pm 2\pi/3$ are being
ignored, except for $\beta$s approaching $\beta_d$ and below, where transitions
between these 3 states are frequent. However, should these be the true vacua,
charge conjugation would be spontaneously broken \cite{DeGrand:2008kx}, and so 
presumably would CP. If so, this could possibly provide a mechanism for 
baryogenesis. 

\section*{Acknowledgements}

DKS is supported in part by US Department of Energy contract DE-AC02-06CH11357.
These simulations were performed on Hopper, Edison and Carver at NERSC, and
Kraken at NICS and Stampede at TACC under XSEDE project TG-MCA99S015,
and Fusion and Blues at LCRC, Argonne. NERSC is supported by DOE contract
DE-AC02-05CH11231. XSEDE is supported by NSF grant ACI-1053575. We thank 
members of the Lattice Higgs Collaboration for informative discussions.


\begin{thebibliography}{99}


\bibitem{Weinberg:1979bn}
  S.~Weinberg,
  Phys.\ Rev.\  D {\bf 19}, 1277 (1979).

\bibitem{Susskind:1978ms}
  L.~Susskind,
  Phys.\ Rev.\  D {\bf 20}, 2619 (1979).


\bibitem{Dietrich:2006cm}
  D.~D.~Dietrich and F.~Sannino,
  Phys.\ Rev.\  D {\bf 75}, 085018 (2007)
  [arXiv:hep-ph/0611341].


\bibitem{Holdom:1981rm}
  B.~Holdom,
  Phys.\ Rev.\  D {\bf 24}, 1441 (1981).

\bibitem{Yamawaki:1985zg}
  K.~Yamawaki, M.~Bando and K.~i.~Matumoto,
  Phys.\ Rev.\ Lett.\  {\bf 56}, 1335 (1986).

\bibitem{Akiba:1985rr}
  T.~Akiba and T.~Yanagida,
  Phys.\ Lett.\  B {\bf 169}, 432 (1986).

\bibitem{Appelquist:1986an}
  T.~W.~Appelquist, D.~Karabali and L.~C.~R.~Wijewardhana,
  Phys.\ Rev.\ Lett.\  {\bf 57}, 957 (1986).


\bibitem{Shamir:2008pb} 
  Y.~Shamir, B.~Svetitsky and T.~DeGrand,
  Phys.\ Rev.\ D {\bf 78}, 031502 (2008)
  [arXiv:0803.1707 [hep-lat]].

\bibitem{DeGrand:2008kx} 
  T.~DeGrand, Y.~Shamir and B.~Svetitsky,
  Phys.\ Rev.\ D {\bf 79}, 034501 (2009)
  [arXiv:0812.1427 [hep-lat]].

\bibitem{DeGrand:2009hu} 
  T.~DeGrand,
  Phys.\ Rev.\ D {\bf 80}, 114507 (2009)
  [arXiv:0910.3072 [hep-lat]].

\bibitem{DeGrand:2010na} 
  T.~DeGrand, Y.~Shamir and B.~Svetitsky,
  Phys.\ Rev.\ D {\bf 82}, 054503 (2010)
  [arXiv:1006.0707 [hep-lat]].

\bibitem{DeGrand:2012yq} 
  T.~DeGrand, Y.~Shamir and B.~Svetitsky,
  Phys.\ Rev.\ D {\bf 87}, 074507 (2013)
  [arXiv:1201.0935 [hep-lat]].

\bibitem{DeGrand:2013uha} 
  T.~DeGrand, Y.~Shamir and B.~Svetitsky,
  Phys.\ Rev.\ D {\bf 88}, 054505 (2013)
  [arXiv:1307.2425 [hep-lat]].


\bibitem{Fodor:2009ar} 
  Z.~Fodor, K.~Holland, J.~Kuti, D.~Nogradi and C.~Schroeder,
  JHEP {\bf 0911}, 103 (2009)
  [arXiv:0908.2466 [hep-lat]].

\bibitem{Fodor:2011tw} 
  Z.~Fodor, K.~Holland, J.~Kuti, D.~Nogradi and C.~Schroeder,
  arXiv:1103.5998 [hep-lat].

\bibitem{Fodor:2012ty} 
  Z.~Fodor, K.~Holland, J.~Kuti, D.~Nogradi, C.~Schroeder and C.~H.~Wong,
  Phys.\ Lett.\ B {\bf 718}, 657 (2012)
  [arXiv:1209.0391 [hep-lat]].

\bibitem{Fodor:2012uw} 
  Z.~Fodor, K.~Holland, J.~Kuti, D.~Nogradi, C.~Schroeder and C.~H.~Wong,
  PoS LATTICE {\bf 2012}, 025 (2012)
  [arXiv:1211.3548 [hep-lat]].

\bibitem{Fodor:2014pqa} 
  Z.~Fodor, K.~Holland, J.~Kuti, D.~Nogradi and C.~H.~Wong,
  PoS LATTICE {\bf 2013}, 062 (2014)
  [arXiv:1401.2176 [hep-lat]].

\bibitem{Fodor:2015vwa} 
  Z.~Fodor, K.~Holland, J.~Kuti, S.~Mondal, D.~Nogradi and C.~H.~Wong,
  PoS LATTICE {\bf 2014}, 244 (2015)
  [arXiv:1502.00028 [hep-lat]].

\bibitem{Fodor:2015eea} 
  Z.~Fodor, K.~Holland, J.~Kuti, S.~Mondal, D.~Nogradi and C.~H.~Wong,
  PoS LATTICE {\bf 2014}, 270 (2015)
  [arXiv:1501.06607 [hep-lat]].

\bibitem{Fodor:2015zna} 
  Z.~Fodor, K.~Holland, J.~Kuti, S.~Mondal, D.~Nogradi and C.~H.~Wong,
  arXiv:1506.06599 [hep-lat].


\bibitem{Hasenfratz:2015ssa} 
  A.~Hasenfratz, Y.~Liu and C.~Y.~H.~Huang,
  arXiv:1507.08260 [hep-lat].


\bibitem{Deuzeman:2008sc} 
  A.~Deuzeman, M.~P.~Lombardo and E.~Pallante,
  Phys.\ Lett.\ B {\bf 670}, 41 (2008)
  [arXiv:0804.2905 [hep-lat]].

\bibitem{Deuzeman:2009mh} 
  A.~Deuzeman, M.~P.~Lombardo and E.~Pallante,
  Phys.\ Rev.\ D {\bf 82}, 074503 (2010)
  [arXiv:0904.4662 [hep-ph]].

\bibitem{Schaich:2012fr} 
  D.~Schaich, A.~Cheng, A.~Hasenfratz and G.~Petropoulos,
  PoS LATTICE {\bf 2012}, 028 (2012)
  [arXiv:1207.7164 [hep-lat]].

\bibitem{Schaich:2015psa} 
  D.~Schaich {\it et al.} [LSD Collaboration],
  arXiv:1506.08791 [hep-lat].


\bibitem{Kogut:2010cz} 
  J.~B.~Kogut and D.~K.~Sinclair,
  Phys.\ Rev.\ D {\bf 81}, 114507 (2010)
  [arXiv:1002.2988 [hep-lat]].

\bibitem{Kogut:2011ty} 
  J.~B.~Kogut and D.~K.~Sinclair,
  Phys.\ Rev.\ D {\bf 84}, 074504 (2011)
  [arXiv:1105.3749 [hep-lat]].


\bibitem{Sinclair:2012fa} 
  D.~K.~Sinclair and J.~B.~Kogut,
  PoS LATTICE {\bf 2012}, 026 (2012)
  [arXiv:1211.0712 [hep-lat]].

\bibitem{Sinclair:2014cga} 
  D.~K.~Sinclair and J.~B.~Kogut,
  PoS LATTICE {\bf 2014}, 239 (2014)
  [arXiv:1410.8494 [hep-lat]].


\bibitem{Lepage:1992xa} 
  G.~P.~Lepage and P.~B.~Mackenzie,
  Phys.\ Rev.\ D {\bf 48}, 2250 (1993)
  [hep-lat/9209022].


\bibitem{Patel:1992vu} 
  A.~Patel and S.~R.~Sharpe,
  Nucl.\ Phys.\ B {\bf 395}, 701 (1993)
  [hep-lat/9210039].

\bibitem{Golterman:1998jj} 
  M.~Golterman,
  Nucl.\ Phys.\ Proc.\ Suppl.\  {\bf 73}, 906 (1999)
  [hep-lat/9809125].


\bibitem{Clark:2006wp}
  M.~A.~Clark and A.~D.~Kennedy,
  Phys.\ Rev.\  D {\bf 75}, 011502 (2007)
  [arXiv:hep-lat/0610047].


\bibitem{Francis:2015lha} 
  A.~Francis, O.~Kaczmarek, M.~Laine, T.~Neuhaus and H.~Ohno,
  Phys.\ Rev.\ D {\bf 91}, no. 9, 096002 (2015)
  [arXiv:1503.05652 [hep-lat]].


\bibitem{Cheng:2011ic} 
  A.~Cheng, A.~Hasenfratz and D.~Schaich,
  Phys.\ Rev.\ D {\bf 85}, 094509 (2012)
  [arXiv:1111.2317 [hep-lat]].

\bibitem{Deuzeman:2012ee} 
  A.~Deuzeman, M.~P.~Lombardo, T.~Nunes Da Silva and E.~Pallante,
  Phys.\ Lett.\ B {\bf 720}, 358 (2013)
  [arXiv:1209.5720 [hep-lat]].


\bibitem{Kogut:1984sb} 
  J.~B.~Kogut, J.~Shigemitsu and D.~K.~Sinclair,
  Phys.\ Lett.\ B {\bf 145}, 239 (1984).

\end{thebibliography}
\end{document}